\documentclass[11pt]{article}

\usepackage{float}
\usepackage[a4paper, total={6in, 8in}]{geometry}
\usepackage{amssymb, amsmath, amsthm}
\usepackage{threeparttable}
\usepackage{booktabs}
\usepackage{setspace}
\usepackage{graphicx}
\usepackage{natbib}
\usepackage[colorlinks=true,linkcolor=black,anchorcolor=black,citecolor=black,filecolor=black,menucolor=black,runcolor=black,urlcolor=black]{hyperref}

\usepackage{bbm}

\usepackage{subcaption}
\usepackage[normalem]{ulem} 
\usepackage{xcolor}         

\begin{document}

\begin{titlepage}
\centering

{\Large\bfseries Market Sensitivities and Growth Differentials Across Australian Housing Markets\par}
\vspace{1em}

{\large Willem P. Sijp\par}
{\normalsize Neoval Pty Ltd \& University of Technology Sydney\par}
\vspace{0.75em}

{\normalsize \today\par} 
\vspace{1.25em}

\begin{abstract}
Australian house prices have risen strongly since the mid-1990s, but growth has been highly uneven across regions. Raw growth figures obscure whether these differences reflect persistent structural trends or cyclical fluctuations. We address this by estimating a three-factor model in levels for regional repeat-sales log price indexes over 1995--2024. The model decomposes each regional index into a national Market factor, two mean-reverting factors (Mining and Lifestyle) that capture decadal cycles that are expressed differently geographically, and a city-specific residual. The Mining factor, proxied by a Perth--Sydney index differential, reflects resource-driven cycles in relative performance; the Lifestyle spread captures amenity-driven coastal and regional pull and the associated city push. The Market loading $\beta_r$ isolates each region's fundamental sensitivity to national growth, so that an assumed national change $f_M$ implies an individual city's growth of $f_r \approx f_M^{\beta_r}$ once mean-reverting spreads are netted out. Comparing realised paths to these factor-implied trajectories indicates when a city is historically elevated or depressed, and attributes the gap to Mining or Lifestyle spreads.

Expanding-window ARIMAX estimation reveals that Market betas are stable across major shocks (the mining boom, the Global Financial Crisis, and COVID-19), while Mining and Lifestyle behave as stationary spreads that widen forecast fans without overturning the cross-sectional ranking implied by $\beta_r$. Once corrected for the geographic cycles, Melbourne amplifies national growth, Sydney tracks the national trend closely, and regional areas dampen it. The framework thus provides a simple, factor-based tool for interpreting regional growth differentials and their persistence.

\end{abstract}

\vspace{0.75em}

\vfill
\end{titlepage}

\setcounter{page}{1}
\onehalfspacing
\doublespacing

\section{Introduction}

Australian house price appreciation has sustained considerable variation across the capital cities and the regional areas over decades. 
  Local housing markets are generally shaped by region-specific conditions, while sharing a common national price trend.
This paper studies such divergence for Australian \emph{houses} (excluding units and apartments) over 1995--2024, using nominal repeat-sales log price indexes. We examine how much of the long-run divergence across cities reflects broadly shared but geographically varying cyclical forces, and how much instead reflects each city's long-run sensitivity to the national market trend. These regional growth differentials matter for housing policy, investment decisions, and spatial inequality: persistent gaps can impede internal migration and exacerbate local skill shortages, while reliable planning requires understanding whether relative growth relationships are stable.

Our approach takes the PCA study of \citet[][]{sijp_francke_ree2024} as a starting point, and separates a trend correlated to the national price index, the Market factor, from two economically interpretable, mean-reverting factors that are expressed differently at various locations. The Mining spread captures the geographic variation in exposure to the resource cycle that raises prices more in mining-exposed regions, relative to the national market and less in Sydney, consistent with its negative Mining factor exposure in both our factor model and the PCA of \citet[][]{sijp_francke_ree2024}. The Lifestyle spread captures amenity-driven cycles in coastal and regional destinations, where markets often lag the Market baseline for extended periods and then catch up sharply during internal migration-driven upswings.

This paper contributes a transparent three-factor decomposition in \emph{levels} that is tightly linked to PCA evidence, yet constructed from simple (weighted) index differences, or ``spreads''. The factors are ``portable'' in the sense that they are reproducible from standard city/region price indexes supplied by multiple vendors. Estimating the model via ARIMAX, we show that these three factors capture most of the low-frequency co-movement and the dominant deviations in Australian city house prices over the last 30 years.

We estimate region-specific factor loadings and document two central findings. First, Market betas $\beta_r$\footnote{the coefficients that multiply the national market factor in the factor model for each city.} are stable in expanding windows, indicating persistent long-run sensitivities to national growth. Second, Mining and Lifestyle behave as stationary factors, allowing the attribution of uncertainty to their long-horizon prediction intervals.

This stability enables transparent scenario mapping: for a multiplicative change $f_M$ in the national price index, the market-only regional change is approximately $f_M^{\beta_r}$, with uncertainty quantified by factor-specific forecast fans \citep{edwards2007}. In the data, Sydney and Canberra (ACT) track the national market closely ($\beta_r\approx 1$), Melbourne and Brisbane amplify national movements, and regional areas dampen them. We emphasize that we do not predict the national path (or even the remaining factors). Instead, we examine city price developments relative to the national index.
While the national Market trend itself may evolve, the relative scaling via $\beta_r$ and the mean-reverting character of the spreads show no evidence of trend instability in expanding windows over 1995--2024, suggesting that the estimated loadings and forecast uncertainties are likely to remain informative in the near future of 5 to 10 years. We do not attempt to predict the national path. Any reference to national growth is a descriptive background rather than a forward-looking projection.

We briefly outline some of the literature on variations in house price growth and underlying drivers. 
Price trajectories can diverge substantially, as evidenced by asynchronous booms across U.S. cities during the 2000s housing cycle \citep{ferreira_gyourko2011}. Individual city growth can exceed the national average over very long periods of time \citep{gyourko_mayer_sinai2013}.
As to the drivers, a long-standing user-cost literature studies the level of house prices relative to rents, interest rates and taxes, typically within an asset-market framework \citep{poterba1984,poterba1992,himmelberg_mayer_sinai2005,campbell_etal2009}. This body of work highlights the role of real mortgage rates, expected capital gains and tax treatment in shaping housing valuations and the rent--price ratio. In the Australian context, empirical studies have related house prices to fundamentals such as income, interest rates and supply constraints \citep[e.g.][]{abelson1994,abelson_etal2005,bodman_crosby2004,otto2007,fox_tulip2014,saunders_tulip2019,abelson_joyeux2023}. \citet{wilson_zurbruegg2008} use cointegration and restriction tests to show that the long-run fundamentals driving house prices differ across Australian capitals. 
\citet{otto2007} documents substantial dispersion in real house price growth and volatility across Australian capitals over 1986--2005, using ABS indexes for established houses. 
He also confirms earlier work by \citet{abelson1994} that Sydney prices tend to lead some of the other cities, which we interpret as evidence of common shocks. In our framework these are captured by a symmetric national Market factor and two geographically structured spread factors (Mining and Lifestyle), so Sydney is just one market loading on shared influences rather than an exogenous driver. We leave the explicit accommodation of leading cities and spillovers within our factor framework for future work.
More generally, our focus and approach differs from the user-cost related literature. Rather than modelling user cost or long-run equilibrium prices as functions of explanatory variables, we take the house price data as a starting point and analyse how a small number of common and geographically structured factors account for differences in long-run price growth across cities.

In examining the last U.S. housing boom, \citet{delnegro2007} employ a Bayesian dynamic factor model in the spirit of \citet{geweke1977} on quarterly OFHEO house price indices for U.S.\ states (1986--2005) to disentangle the common component from the region-specific shocks, finding an increasingly concerted (upward) price movement towards the end of their period (2001--2004). This is an important period as it encompasses the U.S.\ housing boom of the early 2000s. Their returns-based framework isolates a \emph{common} (national) housing factor from local (state- or region-specific) shocks using fixed $\beta$ coefficients for the cities. This establishes two ideas we adopt: first, using a national housing factor to capture broad comovement across cities; and second, using a small set of factors to capture broader geographically varying growth similarities. In contrast to their model, we do not employ a dynamic equation of motion for these factors, but rather base them on time series that are known to explain most of the dataset variance and are associated with broad socio-economic phenomena.

\citet{case2010housing} adopt a risk-related asset-pricing framework, inspired by such models in finance, and model quarterly MSA \emph{returns} against a (differenced) national housing ``market'' factor (proxied by the aggregate U.S.\ house price index), and standard risk characteristics (size/SMB, momentum, and idiosyncratic risk). They find a sizable market effect and, using Fama--MacBeth portfolio tests, evidence that \emph{idiosyncratic} (non-market) risk also helps price housing returns, consistent with a positive risk--return trade-off view a la \citet{fama_french1992}. A related perspective in the Australian context is offered by \citet{morawakage2022}, who, after accounting for spatial dependence and heterogeneity in the Brisbane market, report that at this more granular level, \emph{idiosyncratic} (asset-specific/locational) risks explain excess returns at the submarket (postcode) level, with common speculative forces also playing a role.\footnote{For a broader review of return-based housing asset pricing, see \citet{morawakage2022} and references therein. Our focus is on levels-based growth decomposition rather than the pricing of risk in returns.}

In contrast to these risk-related asset-pricing models formulated in returns, our objective is to characterize persistent relative growth differentials over multi-decadal periods at the aggregated city level, not the cross-sectional pricing of risk. We retain a national market component but operate in levels to preserve long-run co-movement (cf.\ \citealp{holly_etal2010}) and, rather than risk-priced characteristics, we use geographically oriented factors anchored in the PCA evidence of \citet{sijp_francke_ree2024}. Without taking a traded financial asset view \citep{case2010housing} or an alternative user-cost perspective that views owner-occupied housing primarily as a provider of services \citep{himmelberg_mayer_sinai2005}, our factor model is descriptive and remains compatible with either philosophy.

The remainder of the paper proceeds as follows. Section~\ref{sec1} describes the data, the PCA evidence motivating the three-factor structure, and the ARIMAX specification; Section~\ref{sec:results} presents the estimated factor loadings, their stability in expanding windows, and the resulting scenario/uncertainty implications, followed by concluding remarks in Section~\ref{sec:conclusion}.

\section{Methods and Methodology}\label{sec1}

This section outlines the construction of the three-factor model for price indexes of regions within Australia and data used in its empirical implementation.

\subsection{Data}
\label{sec:data}

Our analysis examines house price dynamics across Australian regional markets using transaction records sourced from state valuer general offices and provided by REA Australia. The dataset includes only houses (excluding units and apartments) and spans January 1995 to November 2024, comprising approximately 3 million transactions. Each record contains the sale price in Australian dollars, transaction date, and geospatial coordinates of the property. This data is not used directly. Instead, we construct local price indexes from those records, which serves as input to our three-factor model. As such, our methodology is flexible and can be applied to any set of price indexes.

\paragraph{Geographic hierarchy and index construction.}
We work with three nested levels of geographic aggregation as defined by the Australian Bureau of Statistics \citep{abs:2016}, using the 2016 ASGS (Edition 2) boundaries for SA2/SA3/SA4 and GCCSA regions.
At the finest level, SA2 (Statistical Area Level 2) regions are medium-sized spatial units designed to represent socially and economically cohesive communities, typically containing 3,000–25,000 residents. The 2,052 non-overlapping SA2s provide full national coverage and are roughly comparable to 2.5 U.S. census tracts. These SA2 regions aggregate into 107 SA4 (Statistical Area Level 4) regions, which represent labor markets or clusters of related suburbs. At the coarsest level, we define 15 major GCCSA market areas: the eight state and territory capital cities (Sydney, Melbourne, Brisbane, Perth, Adelaide, Hobart, Darwin, and the Australian Capital Territory) and seven ``Rest of State'' categories encompass all non-metropolitan areas within each state or territory.

Our factor model operates on monthly log price indexes constructed at two levels of geographic aggregation. First, we construct granular SA2-level repeat sales indexes using the spatially regularized Bayesian methods of \citet{sijp_francke_laplacian} as detailed  in~\ref{sec:underlying_indexes}. These 2,052 fine-grained series are used solely for an exploratory principal component analysis (PCA) which guides our selection of interpretable factor proxies. This PCA replicates the approach of \citet{sijp_francke_ree2024}. 

The main analysis, however, is conducted on aggregated indexes at the SA4 and major city levels. These aggregations are formed as weighted averages of the underlying SA2 indexes, with weights proportional to the number of house sales in each SA2 region during January 2015 to January 2020. This weighting scheme ensures that more active markets contribute appropriately to broader regional indexes, while avoiding distortions from transient volume spikes.

Table~\ref{table:counts} summarizes transaction volumes by major market area. Monthly sales per SA2 region range from 2 (ACT, Darwin, Rest of NT) to 7 (Perth), reflecting both market size and the number of constituent SA2 units. Melbourne accounts for the largest share of the sample, with 0.61 million transactions over the full period, followed by Sydney (0.46 million) and Perth (0.32 million). The capital cities collectively account for roughly two-thirds of national house sales, with the remaining third distributed across regional markets.


\begin{table}[ht]
	\centering
\begin{tabular}{lrrr}
\toprule
Market Area & Avg Monthly/SA2 & Monthly Total & Period Total (m) \\
\midrule
Sydney & 5 & 1271 & 0.46 \\
Melbourne & 6 & 1707 & 0.61 \\
Brisbane & 4 & 786 & 0.28 \\
Perth & 7 & 889 & 0.32 \\
Adelaide & 5 & 398 & 0.14 \\
Hobart & 3 & 91 & 0.03 \\
Darwin & 2 & 41 & 0.01 \\
ACT & 2 & 123 & 0.04 \\
Rest of NSW & 5 & 963 & 0.34 \\
Rest of VIC & 6 & 783 & 0.28 \\
Rest of QLD & 4 & 674 & 0.24 \\
Rest of WA & 4 & 254 & 0.09 \\
Rest of SA & 3 & 132 & 0.05 \\
Rest of TAS & 3 & 162 & 0.06 \\
Rest of NT & 2 & 8 & 0.00 \\
\bottomrule
\end{tabular}
    
\caption{Transaction volumes by market area, January 1995–November 2024. Column 2 reports average monthly sales per SA2 region within each market area; Column 3 reports aggregate monthly sales for the market area; Column 4 reports cumulative sales over the 30-year period. ``Rest of'' categories exclude the corresponding capital city. ACT: Australian Capital Territory. State abbreviations: NSW (New South Wales), VIC (Victoria), QLD (Queensland), SA (South Australia), WA (Western Australia), TAS (Tasmania), NT (Northern Territory).}
	\label{table:counts}
\end{table}

As a brief aside, while we report city-level results for interpretability, Sydney exhibits within-city variation. Using SA2 indices (Jan 1995 = 1), the median SA2 has grown $8.83\times$; the interquartile range is $[8.10\times,\,9.86\times]$, with the central 80\% spanning $[7.48\times,\,11.15\times]$. A small right tail (4.3\% of SA2s) exceeds $12\times$, plausibly reflecting localized shocks (such as infrastructure, rezoning, or redevelopment). These pockets widen the forecast bands through the idiosyncratic remainder $\epsilon_r$, but do not overturn $\beta_r$ rankings or our aggregate conclusions.

\subsection{Approximating price indexes using PCA}
\label{sec:pca}

\citet{sijp_francke_ree2024} apply PCA to Australian regional house price indexes and link the three most dominant components to specific socio-economic processes. We replicate their analysis on the SA2 price-index collection described in Section~\ref{sec:underlying_indexes} and recover the same three leading components, which we label Market (PC1), Mining (PC2), and Lifestyle (PC3); their economic interpretation is summarised in Section~\ref{sec:pc_interpretation}. 

As PCA is linear, any regional log price index $\mu_r$ can be approximated by a linear combination of leading principal components:
\begin{equation}
    \mu_r \approx \sum_{k=1}^q a_{kr} z_k,
    \label{eq:z2mu}
\end{equation}
where $q \le p$ denotes the number of retained components and $a_{kr}$ are the associated loadings. This approximation becomes exact when $q=p$. Because the leading components explain most of the variance, truncating at low $q$ (e.g., $q=2$ or $3$) already yields a strong representation. Conversely, orthogonality implies each component can be expressed as a weighted sum of regional indexes:
\begin{equation}
    z_k = \sum_{r=1}^p a_{kr} \mu_r .
    \label{eq:mu2z}
\end{equation}
This bidirectional relationship motivates a parsimonious factor model in which transparent index-based proxies for $z_1$, $z_2$, and $z_3$ capture the essential co-movement structure while remaining stable across expanding estimation windows.

\subsection{Economic interpretation of the leading PCs}
\label{sec:pc_interpretation}

We briefly summarise the economic interpretation of the first three PCs estimated here, originally documented in \citet{sijp_francke_ree2024}. 

PC1 (``Market'') is a broad national component capturing the common drift in dwelling values across population centres. Regions with higher loadings on PC1 tend to move more than one-for-one with aggregate Australian housing cycles. We verify that the first component $z_1$ tracks the national price index $U$ (the weighted mean of SA2 indexes) almost identically, with correlation exceeding 0.99 (Table~\ref{table:corr_pca}).

PC2 (``Mining'') reflects resource-sector dynamics. Correlations with mining investment \citep[calculated in][]{sijp_francke_ree2024} indicate that PC2 is most strongly expressed in resource-exposed regions (e.g.\ WA, QLD). Its time profile tracks the 2000s mining-investment boom and its aftermath, consistent with major macro shocks over that period \citep{downes_et_al_rba_report2014}. Labor migration during the mining boom \citep{foo_salim_minecon2022} likely contributed to these regional price pressures. Loadings on this component are negative in Sydney (and to a lesser extent Melbourne), indicating that, relative to mining-exposed regions, these markets exhibit weaker or opposite-signed comovement with the resource cycle. See Section~\ref{sec:examples_indexes} for further discussion.

PC3 (``Lifestyle'') loads positively on amenity-rich coastal and regional areas associated with ``sea/tree-change'' migration rather than employment centres. This reflects a search for a more affordable lifestyle, often in scenic locations, with a better work-life balance, increasingly enabled by remote work.
Evidence more generally documents households moving from capital cities toward regional towns, including lifestyle destinations \citep{bohnet2010, yanotti2023}. This has intensified during and after COVID-19, but the pattern is evident for Lifestyle regions throughout the last 30 years, with the penultimate upswing in 2000-2004. These regions lag capitals during booms, then catch up rapidly near the end, experiencing affordability pressure as migration inflows (from the cities) rise \citep{sijp2025report}.
This movement also agrees with the common knowledge of the long-standing pattern of households moving from the colder southern states to the warmer state of Queensland and coastal regions of NSW.
Among the evidence to link PC3 to lifestyle is the correlation between PC3 loadings and word frequencies in real estate advertisement listings: high-PC3 regions show elevated frequencies of lifestyle descriptors (e.g., \emph{beach}, \emph{coastal}, \emph{lifestyle}).\footnote{See \citet{sijp2025report} for the word-count study linking listing text to PC3.}  Note that these loadings are produced by the PCA, they are not a-priori identified with mining and lifestyle.


\subsection{Identifying the factors for the factor model}
\label{sec:identifying_factors}

The linear approximation of regional indexes by PCA components (Equation~\ref{eq:z2mu}) motivates a factor model in which each $\mu_r$ is expressed as a linear combination of a small number of interpretable factors plus an autocorrelated residual. 
Our aim is to construct such a model using simple, readily available factors derived from aggregate price indexes that proxy the leading principal components while remaining stable under expanding estimation windows. 
To this end, we construct the non-trend factors in terms of concrete differences in broad indexes between locations that replicate the PC behavior but are grounded in intuitive price differentials and are portable: the Market index and Mining/Lifestyle differences can be reproduced directly from standard city or regional price indexes available from multiple data vendors, making the framework easy to implement on other datasets without the need for a large underlying national collection of granular indexes. 

For clarity, we observe that the PCA derived series $z_1$, $z_2$ and $z_3$ correspond directly to the first three principle components: PC1 (Market), PC2 (Mining), PC3 (Lifestyle) respectively. The factor proxies introduced below are constructed explicitly to serve as empirically stable approximations to these components.

\paragraph{Factor 1: Market.}
 PC1, $z_1$, is nearly identical to the national price index $U$, we therefore adopt $U$ as our Market factor. This represents the common trend across all regions and absorbs the unit-root component of the system. Here, we have used the PCA procedure to guide our choice of market factor, and the PCA is independent of the weights used to construct a national index: it is solely based on the co-movement of the regions irrespective of weight.

\begin{figure}[H]
    \begin{center}
    \scalebox{1}{\includegraphics[width=0.95\linewidth]{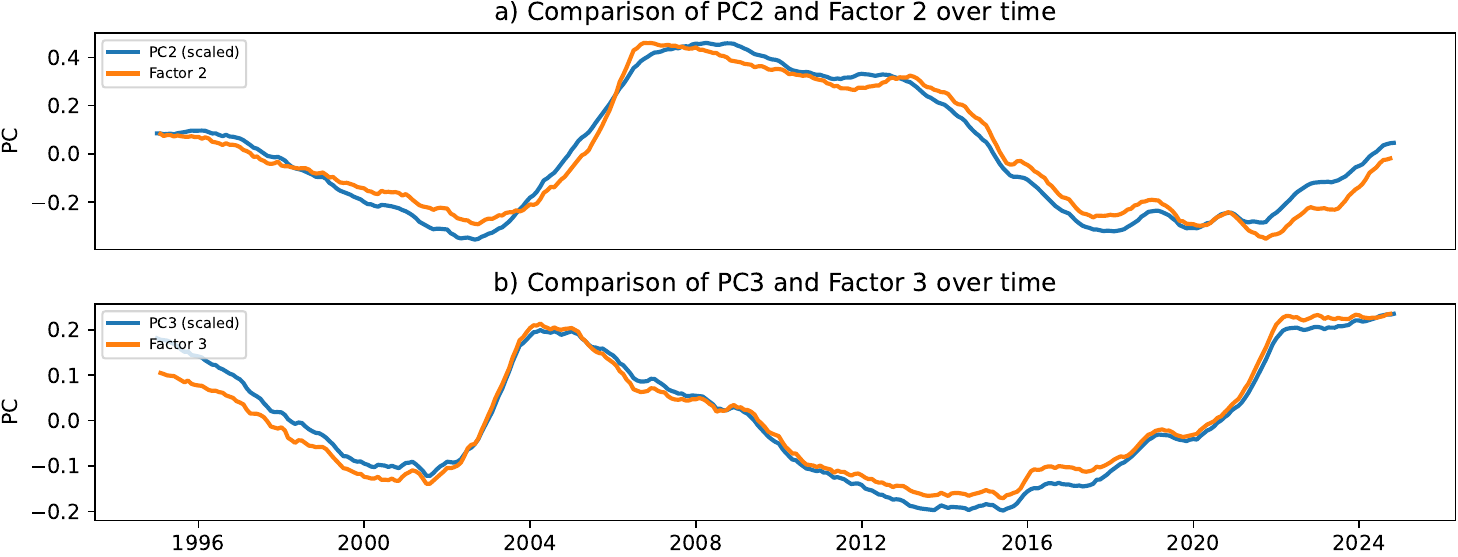}}
    \end{center}
    \caption{Comparison of PC series 2 (Mining vs Sydney) and 3 (Lifestyle) to their factor approximations: (a) the Perth--Sydney spread $\delta_{PS} = \mu_{P} - \alpha \mu_{S}$ alongside PC2, and (b) the Lifestyle spread alongside PC3.}
    \label{fig:compare_factor_pca}
\end{figure}

\paragraph{Factor 2: Mining (Perth--Sydney spread).}
\citet{sijp_francke_ree2024} show that PC2 captures opposing dynamics between mining-exposed regions (positive loadings) and Sydney (negative loading). To proxy $z_2$, we construct the Perth--Sydney spread
\[
\delta_{PS} = \mu_P - \alpha \mu_S,
\]
where $\mu_P$ and $\mu_S$ are the log indexes for Greater Perth and Greater Sydney, and $\alpha \approx 1$ is a trend-adjustment coefficient chosen to filter out the common national trend. We estimate $\alpha$ by
\[
\alpha = \frac{\text{cov}(\mu_P, U)}{\text{cov}(\mu_S, U)},
\]
yielding $\alpha = 0.94$ for our data. This ratio ensures that the spread $\delta_{PS}$ has a low correlation with the Market factor $U$, isolating the Perth-Sydney differential net of common national growth. Among the candidate bilateral spreads evaluated, the Perth–Sydney differential demonstrated the strongest empirical correspondence with PC2. Perth's economy is heavily resource-dependent while Sydney's is services-dominated and has the most negative PCA 2 loadings, rendering their relative performance a natural proxy for resource-sector cycles.
Figure~\ref{fig:compare_factor_pca}a compares $\delta_{PS}$ to PC2; the two series track each other closely, with a correlation of 0.98 (Table~\ref{table:corr_pca}).


\paragraph{Factor 3: Lifestyle.}
For PC3, there is no single city-level spread that naturally proxies the component. Instead, we construct a Lifestyle spread by taking an average weighted by the number of houses (using the weights defined in~\ref{sec:underlying_indexes}) of the top 20 lifestyle SA4 regions according to their PC3 loadings, and offsetting this against the bottom 20 regions. This isolates the strongest lifestyle signal while avoiding noise from weakly loaded regions. \footnote{\textbf{Top Lifestyle regions (positive PC3 loadings):} Central Coast, Hunter Valley Exc Newcastle, South Australia - South East, Ipswich, Capital Region, Hobart, Central West, South East, Moreton Bay - North, Riverina, Southern Highlands And Shoalhaven, Richmond - Tweed, Coffs Harbour - Grafton, Murray, Gold Coast, West And North West, Launceston And North East, Sunshine Coast, Wide Bay, Mid North Coast. \textbf{Bottom Lifestyle regions (negative PC3 loadings):} Sydney Inner West, Western Australia - Wheat Belt, Sydney City And Inner South, Perth - Inner, Melbourne - Outer East, Perth - North West, Melbourne - South East, Western Australia - Outback (south), Perth - South East, Perth - North East, Melbourne - North West, Melbourne - North East, Melbourne - West, Melbourne - Inner South, Melbourne - Inner East, Melbourne - Inner, Western Australia - Outback (north), Queensland - Outback, Darwin, Northern Territory - Outback.} 
As with the Mining spread, we apply a trend-adjustment coefficient $\alpha = 0.82$ to remove the common market factor. Figure~\ref{fig:compare_factor_pca}b shows the Lifestyle spread alongside PC3; the two series track closely, with a correlation of 0.98 (Table~\ref{table:corr_pca}). 


\begin{table}[ht!]
	\centering
    \small
\begin{tabular}{rlll}
\toprule
PC series & Factor & Corr & Description \\
\midrule
1 & National Index & 1.00 & Market \\
2 & Perth-Sydney Spread & 0.98 & Mining vs Sydney \\
3 & Lifestyle Spread & 0.98 & Lifestyle \\
\bottomrule
\end{tabular}
 \caption{Pairwise correlations between the PCA components and the corresponding factors. }
    \label{table:corr_pca}
\end{table}

Table~\ref{table:corr_pca} summarizes the correspondence between PCA components and our three factors, with a close correlation to their PCA counterparts for all factors, particularly the Market factor, that is effectively identical to PC1 (correlation 1.00). This close agreement validates our factor proxies: they capture the same co-movement structure as PCA.

Finally, Table~\ref{table:cov} shows that the three factors are nearly orthogonal, with pairwise correlations between -0.18 and 0.19. This near-independence simplifies estimation and interpretation by ensuring that each factor contributes additively to regional price dynamics without strong multicollinearity, and remains consistent with the orthogonality of the underlying principal components. Note that the trend‐adjustment coefficients $\alpha$ were chosen based on criteria independent to the model selection presented in later sections, and were held fixed throughout all specifications. 


\begin{table}[ht!]
	\centering
    \small
\begin{tabular}{lrrr}
\toprule
 & market & mining & lifestyle \\
\midrule
market & 1.00 & -0.06 & 0.19 \\
mining & -0.06 & 1.00 & -0.18 \\
lifestyle & 0.19 & -0.18 & 1.00 \\
\bottomrule
\end{tabular}
 \caption{Correlations between the three factors. The factors are near-orthogonal.}
    \label{table:cov}
\end{table}

\subsection{The factor model and uncertainty in projections relative to the market}
\label{sec:model}

The model decomposes each regional index $\mu_r$ into a Market component (capturing long-run growth), a Mining component (capturing resource-sector cycles), and a Lifestyle component (capturing amenity-driven demand shifts), plus a city-specific idiosyncratic remainder. The Market loadings determine long-run growth differentials, while the remaining factors are thought of as bounded, representing socio-economic cycles.

More concretely, we model the (log) regional index $\mu_r$ as
\begin{equation}
    \mu_r(t) \;=\; b_r \;+\; \beta_r\,U(t) \;+\; \lambda_r\,\delta_{PS}(t) \;+\; \gamma_r\,\delta_{L}(t) \;+\; \epsilon_r(t),
    \label{eq:main}
\end{equation}
where $U$ is the national market index, $\delta_{PS}$ the Perth--Sydney spread, $\delta_{L}$ the Lifestyle spread, and $\epsilon_r$ an idiosyncratic disturbance. Although autocorrelated, we assume $\mathbb{E}[\delta_{PS}]=\mathbb{E}[\delta_{L}]=\mathbb{E}[\epsilon_r]=0$.
We treat $(\beta_r,\lambda_r,\gamma_r)$ as time-invariant constants, providing justification for this in Section~\ref{sec:stability}. The coefficients $(\beta_r,\lambda_r,\gamma_r)$ represent regional sensitivities to the three factors, analogous to loadings on the corresponding principle components. 

Our market–only projection at $t_2$ is $\widehat{\mu}_r(t_2)=\beta_r\,U(t_2)$. We are interested in performance \emph{relative} to the national market. Over a horizon $[t_1,t_2]$, let price growths for a specific location resp. the national market be:
\[
f_r \;=\; e^{\mu_r(t_2)-\mu_r(t_1)},\qquad
f_M \;=\; e^{U(t_2)-U(t_1)}.
\]
(So a price $p_1$ at $t_1$ simply grows to $p_2=f_rp_1$ at $t_2$.) For instance, if the national index doubles ($f_M=2$), the market–only proxy implies $f_r \approx 2^{\beta_r}$; $\beta_r>1$ amplifies national moves, $\beta_r<1$ dampens them. \footnote{The coefficient $\beta_r$ reflects the responsiveness of regional price growth with respect to national market movements. The interpretation holds under the assumption of log-linear growth and continuous compounding.}

To quantify the uncertainty of this approximation, we define the log-multiplicative error over the horizon $[t_1,t_2]$ as:
\[
\ell_r(t_1,t_2)
= \log\!\left(\frac{e^{\beta_r\,[U(t_2)-U(t_1)]}}{e^{\mu_r(t_2)-\mu_r(t_1)}}\right)
= -\,\lambda_r\,[\delta_{PS}(t_2)-\delta_{PS}(t_1)]
  -\,\gamma_r\,[\delta_{L}(t_2)-\delta_{L}(t_1)]
  -\,[\epsilon_r(t_2)-\epsilon_r(t_1)].
\]
Let $\mathcal{H}_{t_1}$ be the information set at $t_1$ and define the horizon length as $h=t_2-t_1$. 
Since $X(t_1)$ is known at $t_1$ for a series $X$, the variance of the level change equals the $h$-step level forecast variance: $
\mathrm{Var}\!\big(X(t_2)-X(t_1)\,\big|\,\mathcal{H}_{t_1}\big)
\;=\; \mathrm{Var}\!\big(X(t_2)\,\big|\,\mathcal{H}_{t_1}\big).
$
We therefore define the three horizon variances (``fans'') as: 
\[
\sigma^2_{PS}(h) \;:=\; \mathrm{Var}\!\big(\delta_{PS}(t_2)\mid\mathcal{H}_{t_1}\big),\qquad
\sigma^2_{L}(h) \;:=\; \mathrm{Var}\!\big(\delta_{L}(t_2)\mid\mathcal{H}_{t_1}\big),\qquad
\sigma^2_{\epsilon}(h) \;:=\; \mathrm{Var}\!\big(\epsilon_r(t_2)\mid\mathcal{H}_{t_1}\big).
\]
Assuming the three components are approximately orthogonal over the horizon, the variance of the log error is: 
\[
\mathrm{Var}\!\big[\ell_r(t_1,t_2)\mid\mathcal{H}_{t_1}\big]
\;\approx\;
\lambda_r^{2}\,\sigma^2_{PS}(h)\;+\;\gamma_r^{2}\,\sigma^2_{L}(h)\;+\;\sigma^2_{\epsilon}(h).
\]
A corresponding $95\%$ \emph{multiplicative} uncertainty band for the market–only growth approximation
$f_r \approx f_M^{\beta_r}$ is:
\begin{equation}
    e^{\pm\,1.96\,\sqrt{\lambda_r^{2}\,\sigma^2_{PS}(h)+\gamma_r^{2}\,\sigma^2_{L}(h)+\sigma^2_{\epsilon}(h)})}.
    \label{eq:uncertainty}
\end{equation}

Unlike many regressions of returns in financial time series, the city-specific idiosyncratic disturbance $\varepsilon_{r,t}$ is autocorrelated (this would even be the case if we were to model housing index returns). Therefore, we do not estimate Eq.~\ref{eq:main} with OLS. Instead, we model the disturbance as a low-order ARMA process,
\begin{equation}
  \varepsilon_{r,t}
  = \sum_{i=1}^{p_r} \phi_{r,i}\,\varepsilon_{r,t-i}
    + u_{r,t}
    + \sum_{j=1}^{q_r} \theta_{r,j}\,u_{r,t-j},
  \qquad u_{r,t} \sim \text{i.i.d. }(0,\sigma_r^2),
  \label{eq:arimax_error}
\end{equation}
so that \eqref{eq:main} and \eqref{eq:arimax_error} define a standard ARIMAX$(p_r,0,q_r)$ dynamic regression: the city index $\mu_r(t)$ is regressed on the contemporaneous Market, Mining and Lifestyle factors, while the disturbance follows an ARMA$(p_r,q_r)$ process. (Equivalently, this can be written in an autoregressive distributed-lag form in $\mu_r(t)$ and the factors.) We denote by $e_{r,t}$ the residuals from the fitted ARIMAX model (observed index minus fitted value).

The regional log real price indices $\mu_r(t)$ and the Market factor $U_t$ are nonstationary, whereas the Mining and Lifestyle spreads are approximately stationary (\ref{app:mining_regimes}), so we treat the factor regression in levels as describing a long-run relationship between $\mu_r(t)$ and $U_t$, with the bounded dynamics captured by the remaining factors an $\epsilon$. In Section~\ref{sec:examples_indexes} we show that the $\epsilon$ are bounded and display no visible trend, which is consistent with an approximately stationary process.

\noindent\textit{Remark.}  The quantities $\sigma^2_{PS}(h)$, $\sigma^2_{L}(h)$, and $\sigma^2_{\epsilon}(h)$ are the $h$-step conditional \emph{level} variances (forecast “fans”) implied by the fitted dynamics of the spreads and the idiosyncratic component; loadings are fixed and are not projected.

\section{Results}
\label{sec:results}

City price indexes are modeled with three-factors and a city-specific component ($\epsilon$).
A concern of this paper is the temporal stability of the factor loadings, particularly $\beta_r$, i.e. with respect to an expanding fit window. In addition, to assess the uncertainties and risks introduced by the remaining mean-reverting factors, we are interested in their forecast fans over a future period: we estimate this by individual ARIMA models for each factor. Finally, We adopt an ARIMAX specification for the factor model of each individual city in Equation~\ref{eq:main}, allowing an estimation of the regression coefficients and the construction of a forecast funnel for the autocorrelated remainder.

\subsection{Factor model selection for cities}\label{sec:model_selection}

We estimate ARIMAX$(p,0,q)$ models for Equation~\ref{eq:main} in levels ($d=0$), with the unit-root absorbed by the Market factor $U_t$; Mining and Lifestyle enter as spreads, and the remainder $\epsilon_r(t)$ is treated as stationary.
Although an ADF test on the raw Mining spread does not reject a unit-root, this mainly reflects structural breaks around the mining boom; \ref{app:mining_regimes} shows that once break regimes are accounted for, the Mining factor is well described as a persistent but mean-reverting cycle, consistent with our levels ARIMAX specification.

We will see that visual inspection shows no drift in $\epsilon_r(t)$ after accounting for the factors (Section~\ref{sec:examples_indexes}, Figures~\ref{fig:factor_series_rest_nsw}--\ref{fig:factor_series_perth}).
Non-seasonal orders $(p,q)\in\{0,1,2\} \times \{0,1,2\}$ are chosen by AICc \citep{hurvich_tsai1989} with a seasonal MA(1) at $s=12$; low orders dominate (typically $(2,0,1)$ or $(2,0,0)$). Full selections and Ljung--Box diagnostics \citep{ljung_box1978} are reported in Appendix Table~\ref{app:orders}.
LB12 (and LB24) denote Ljung–Box portmanteau $p$-values testing for no residual autocorrelation up to lag 12 (24), computed on the one-step-ahead prediction errors $e_r(t)=y_r(t)-\hat y_r(t\mid t-1)$ from each fitted ARIMAX (unstandardized). Degrees of freedom are not adjusted for estimated ARMA/seasonal parameters.

\paragraph{Lifestyle inclusion}
We wish to include at least two factors, and now assess whether adding Lifestyle improves fit by comparing Market+Mining (2f) to Market+Mining+Lifestyle (3f). Inclusion is favored when $\Delta\mathrm{AICc}\le -2$ \citep{burnham_anderson2002}. Ten of 13 regions meet this rule, with large gains for Melbourne ($-287.5$), Rest of QLD ($-100.2$), and Rest of NSW ($-98.3$). Residual diagnostics on $e_{r,t}$ are generally satisfactory (LB12) with some exceptions, so we adopt the three-factor specification across the board for comparability and motivated by the PCA; see Table~\ref{table:lifestyle_inclusion}. For example, although the Rest of QLD  does not pass the LB12 test under 3f (unlike 2f), we nonetheless see that $\Delta\mathrm{AICc}$ provides a large improvement. Residual variance also shrinks (not shown).


\begin{table}[ht!]
\centering
\small
\begin{tabular}{lrrrrrrl}
City & AICc 2f & AICc 3f & dAICc & LB12 2f & LB12 3f & Better \\
Melbourne & -2952.33 & -3239.83 & -287.50 & 0.49 & 0.64 & Yes$^{\ast}$ \\
Rest Of QLD & -2928.69 & -3028.90 & -100.21 & 0.73 & 0.00 & Yes \\
Rest Of NSW & -3040.94 & -3139.19 & -98.25 & 0.22 & 0.10 & Yes$^{\ast}$ \\
Brisbane & -2912.16 & -2974.46 & -62.30 & 1.00 & 0.88 & Yes$^{\ast}$ \\
Hobart & -2210.43 & -2250.70 & -40.27 & 0.17 & 0.09 & Yes$^{\ast}$ \\
Sydney & -3143.69 & -3161.14 & -17.45 & 0.56 & 0.08 & Yes$^{\ast}$ \\
Perth & -3185.99 & -3202.31 & -16.32 & 0.55 & 0.09 & Yes$^{\ast}$ \\
Rest Of WA & -2568.58 & -2582.01 & -13.43 & 0.01 & 0.04 & Yes \\
Rest Of SA & -2732.60 & -2736.21 & -3.61 & 0.19 & 0.23 & Yes$^{\ast}$ \\
Rest Of VIC & -2912.95 & -2915.59 & -2.64 & 0.85 & 0.95 & Yes$^{\ast}$ \\
Adelaide & -2760.06 & -2762.06 & -1.99 & 0.14 & 0.14 & No \\
Darwin & -2283.53 & -2285.36 & -1.83 & 1.00 & 0.99 & No \\
ACT & -2284.58 & -2283.96 & 0.62 & 0.54 & 0.49 & No \\
\end{tabular}
\caption{Two– vs three–factor ARIMAX in levels (Market+Mining vs Market+Mining+Lifestyle). 
Reported are AICc for each spec and $\Delta\mathrm{AICc}=\mathrm{AICc}_{\text{3f}}-\mathrm{AICc}_{\text{2f}}$ (negative favors three factors). 
Lifestyle is included when $\Delta\mathrm{AICc}\le -2$; The asterisk in “Yes$^{\ast}$” indicates LB12 $\ge 0.05$.}
\label{table:lifestyle_inclusion}
\end{table}

\subsection{The stability of the factor series}
\label{sec:stability_factors}

\paragraph{The factor-specific ARIMA models and the uncertainties from Mining and Lifestyle.}
To estimate the uncertainties of Equation~\ref{eq:uncertainty} in Section~\ref{sec:model}, we need to first obtain the forecast uncertainties of the individual Mining and Lifestyle factor time series (as our work is relative to the market, we do not need this for $U$): to this end, we employ an individual ARIMA model for each of the two factors\footnote{This is a separate model for each of the 2 factors, not to be confused with the city-specific ARIMAX estimations of the factor model.}.

For each factor we grid–search ARIMA$(p,d,q)$\footnote{Grid: $p\in\{0,1,2,3\}$, $q\in\{0,1,2\}$, $d\in\{0,1\}$. }
 (intercept only; no seasonals), using SARIMAX. Models are ranked by AICc and we apply a parsimony rule: among specifications within $\Delta$AICc $\le 2$ of the best at a given $d$, we pick the simplest (minimal $p{+}q$). We then prefer $d{=}0$ unless the best $d{=}1$ improves AICc by a large margin (10 points). Under this procedure both spreads select ARIMA$(2,0,1)$, and residual Ljung–Box tests at lags 12 and 24 remain high (Table~\ref{table:factor_funnels}). We prefer to model Mining and Lifestyle as stationary (or, if selected, highly persistent) cycle indices, consistent with their origin as PC2/PC3 components where PC1 (Market) absorbs the dominant long-run trend—so strong drift in these spreads is not expected.
The coefficients are reported in \ref{app:arima_factors}.


\begin{figure}[H]
    \begin{center}
    \scalebox{1}{\includegraphics[width=0.95\linewidth]{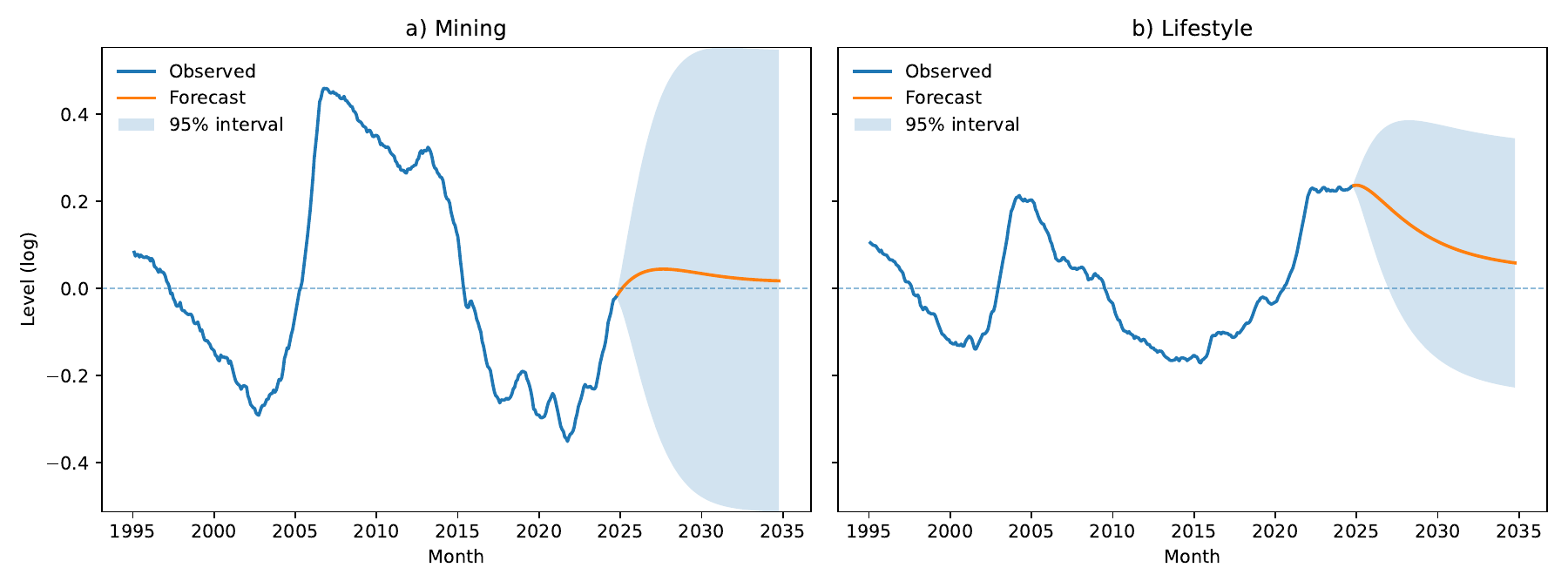}}
    \end{center}
\caption{Ten-year ARIMA forecasts and 95\% prediction intervals (“fans”) for the factor spreads: (a) Mining; (b) Lifestyle. }
   \caption*{\footnotesize  \textit{Notes}: Black lines show the observed series; solid lines show the forecast means; shaded bands are conditional 95\% state-space intervals. Models are estimated in levels with orders chosen by AICc using a \(\Delta\)AICc=2 parsimony rule; (2,0,1) for both factors.  }
    \label{fig:factor_funnels}
\end{figure}

The ten-year ARIMA forecasts and 95\% prediction intervals for the factor spreads shown in Figure~\ref{fig:factor_funnels} demonstrate that both spread series remain reasonably bounded in the future. We use these ARIMA-based fans as inputs to the uncertainty estimates in Equation~\ref{eq:uncertainty}, interpreting them as conditional on the maintained assumption that Mining and Lifestyle are stationary (or highly persistent) spreads.

\paragraph{Magnitudes at the 10–year horizon.}
Table~\ref{table:factor_funnels} reports the implied 120–month–ahead standard deviations and their multiplicative 95\% factors at the \emph{factor level}. For Mining we obtain \(\sigma_{10y}=0.27\), corresponding to a factor–level band of \(x95=\exp(1.96\,\sigma)=1.70\); Lifestyle is tighter with \(\sigma_{10y}=0.15\) and \(x95=1.33\). In multiplicative terms, a 70\% range for Mining and 33\% range for Lifestyle at the ten-year horizon indicate moderate long-run uncertainty that we can use in our city-specific estimations. Importantly, these values are factor-level uncertainties and must be scaled by regional loadings to obtain region-specific bands. In our regional scenario projections, the factor contributions enter quadratically via
\[
\lambda_r^2\,\sigma^2_{PS}(h)\quad\text{and}\quad \gamma_r^2\,\sigma^2_{L}(h),
\]
so regions with larger \(|\lambda_r|\) (resp. \(|\gamma_r|\)) inherit proportionally more Mining (resp. Lifestyle) uncertainty. The idiosyncratic remainder \(\sigma^2_{\epsilon}(h)\) is then added to form the total variance \(\sigma^2_{\text{total}}(h)\) used in the regional bands reported in the next section (Table~\ref{table:factor_coefs}).


\begin{table}[ht!]
	\centering
    \small

\begin{tabular}{lrrrrrll}
\toprule
Factor & Mining & Lifestyle \\
\midrule
ARIMA$(p,0,q)$ & (2,0,1) & (2,0,1) \\
AICc & -2542.32 & -2887.47 \\
LB12 & 0.84 & 0.90 \\
LB24 & 0.93 & 0.99 \\
$\sigma$ (10y) & 0.27 & 0.15 \\
$x95$ (factor) & 1.70 & 1.33 \\
\bottomrule
\end{tabular}
 \caption{ARIMA$(p,0,q)$ fits in levels with intercept (no seasonals) for the Mining and Lifestyle factor processes. Reported are AICc and Ljung--Box $p$-values at lags 12 and 24 (residual whiteness). $\sigma$ (10y) is the 120-month ahead standard deviation of the factor forecast; $x95$ (factor) $=\exp(1.96,\sigma)$ is the corresponding factor-level 95\% multiplicative band.   }

    \label{table:factor_funnels}
\end{table}

The stable fans and whitened residuals validate our use of fixed loadings with uncertainty propagated through factor–specific forecast variances. 

\subsection{The factor loadings and their stability}
\label{sec:stability}

\begin{figure}[H]
  \centering

  \begin{subfigure}[t]{0.49\linewidth}
    \caption{Greater Sydney}
    \vspace{0.25ex}
    \includegraphics[width=\linewidth]{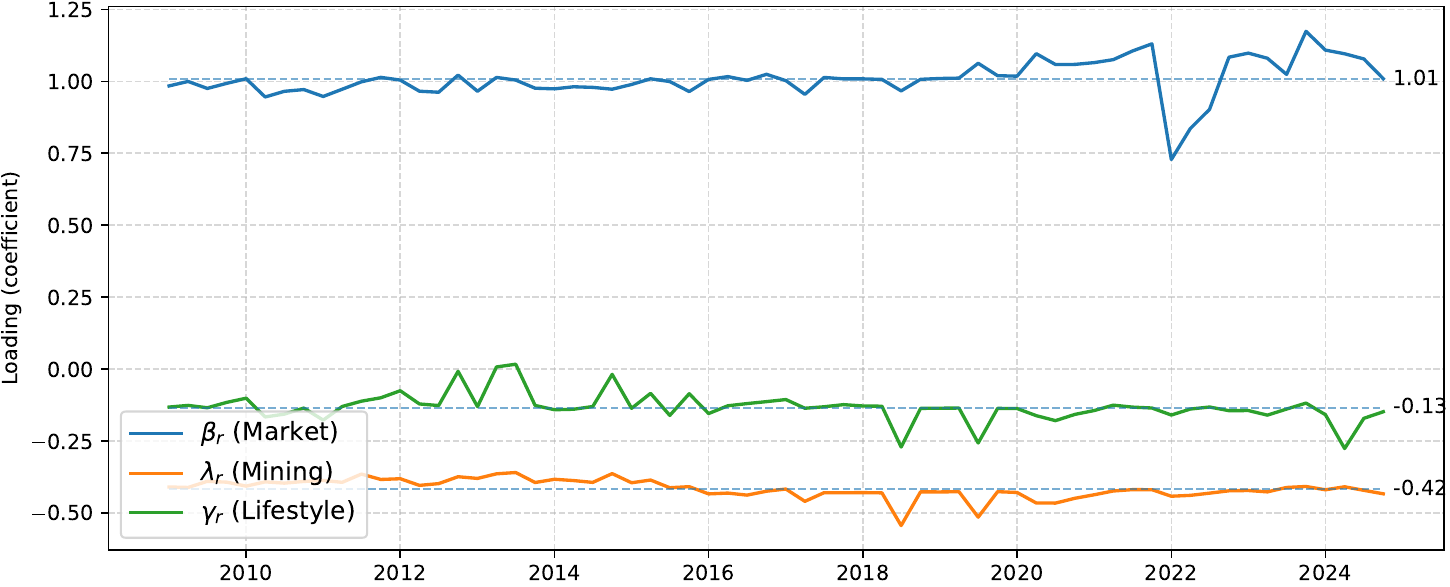}
  \end{subfigure}
  \begin{subfigure}[t]{0.49\linewidth}
    \caption{Greater Melbourne}
    \vspace{0.25ex}
    \includegraphics[width=\linewidth]{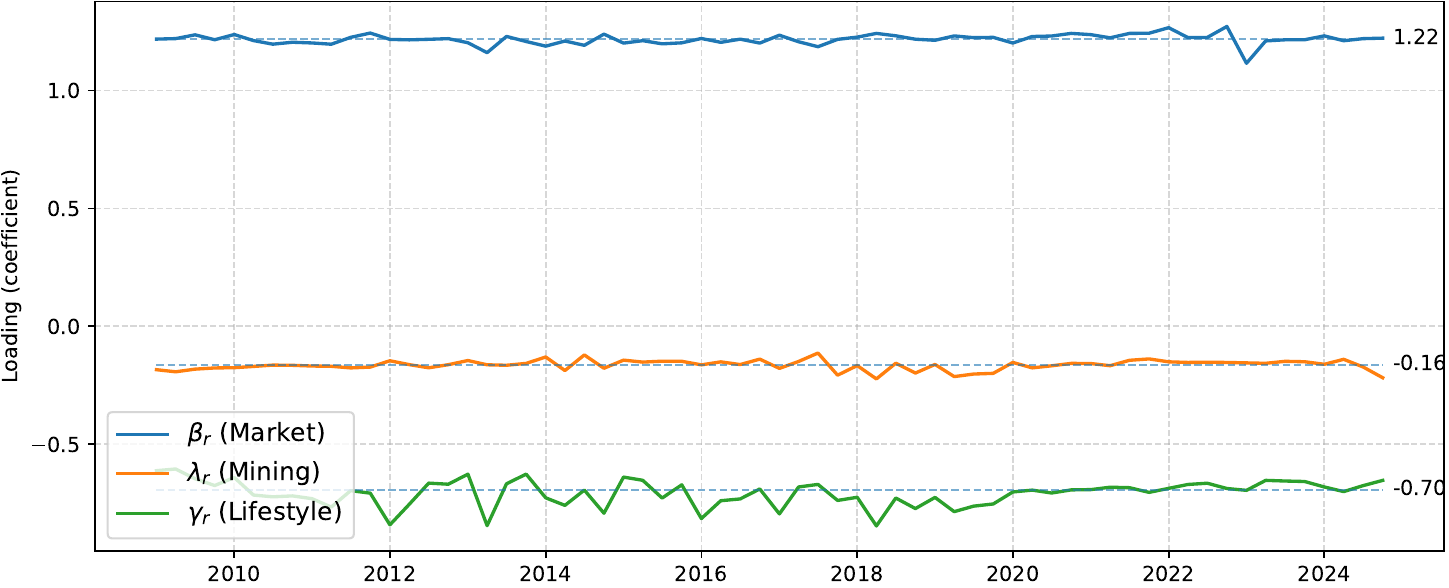}
  \end{subfigure}

  \medskip

  \begin{subfigure}[t]{0.49\linewidth}
    \caption{Greater Brisbane}
    \vspace{0.25ex}
    \includegraphics[width=\linewidth]{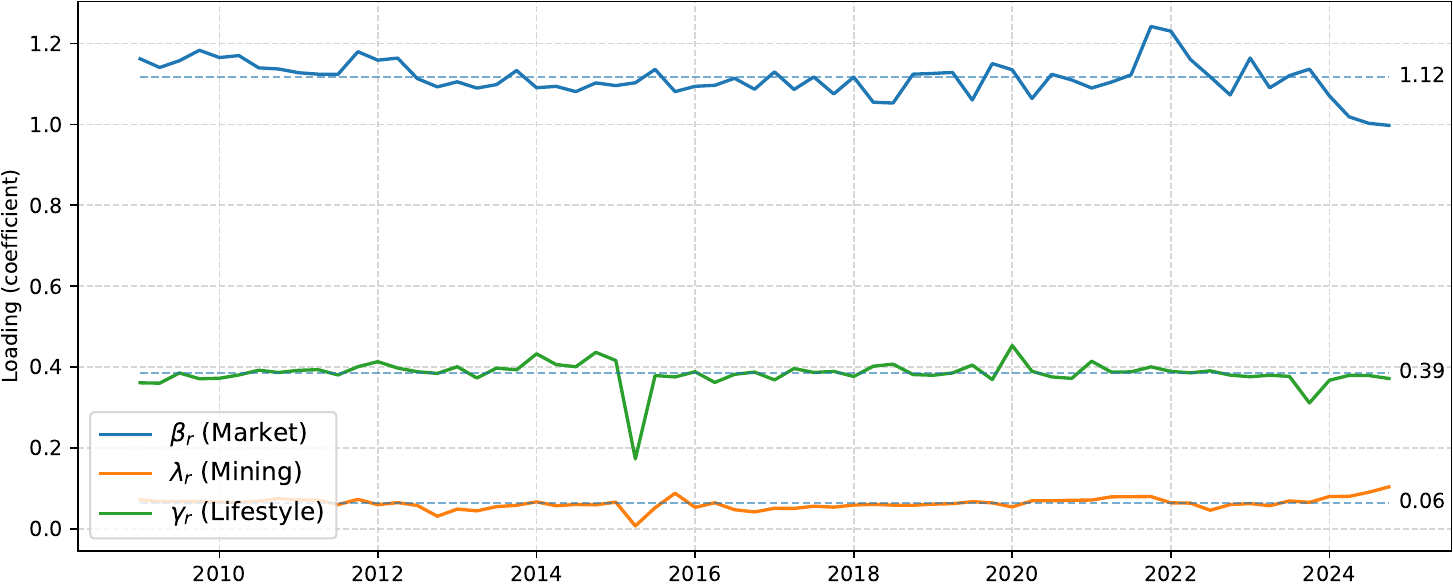}
  \end{subfigure}
  \begin{subfigure}[t]{0.49\linewidth}
    \caption{Greater Perth}
    \vspace{0.25ex}
    \includegraphics[width=\linewidth]{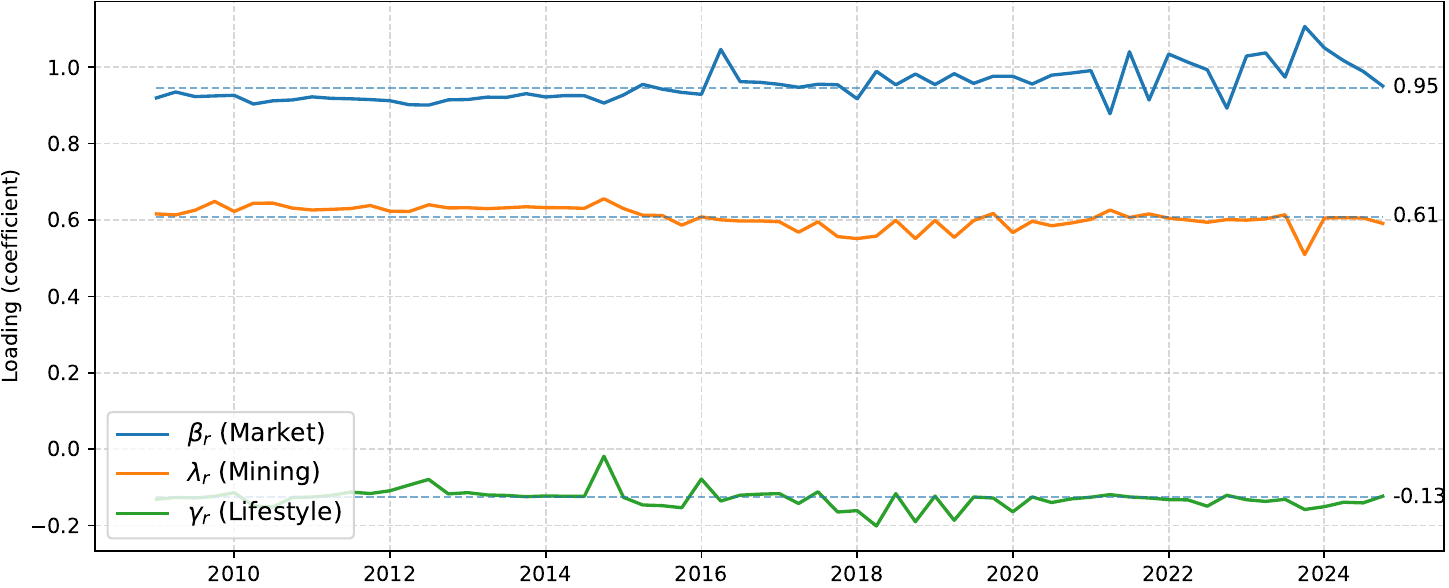}
  \end{subfigure}

  \caption{Expanding-window ARIMAX with seasonal MA(1) (\(s=12\)) factor loadings for major cities. All windows start in January 1995; the endpoint advances from January 2008 to December 2024 in three-month steps. Factor loadings are colored as: Market (\(\beta_r\)): blue; Mining (\(\lambda_r\)): orange; Lifestyle (\(\gamma_r\)): green. }
  \label{fig:expand_panel_cities_topcaps}
\end{figure}

\begin{figure}[H]
  \centering

  \begin{subfigure}[t]{0.49\linewidth}
    \caption{Rest of NSW}
    \vspace{0.25ex}
    \includegraphics[width=\linewidth]{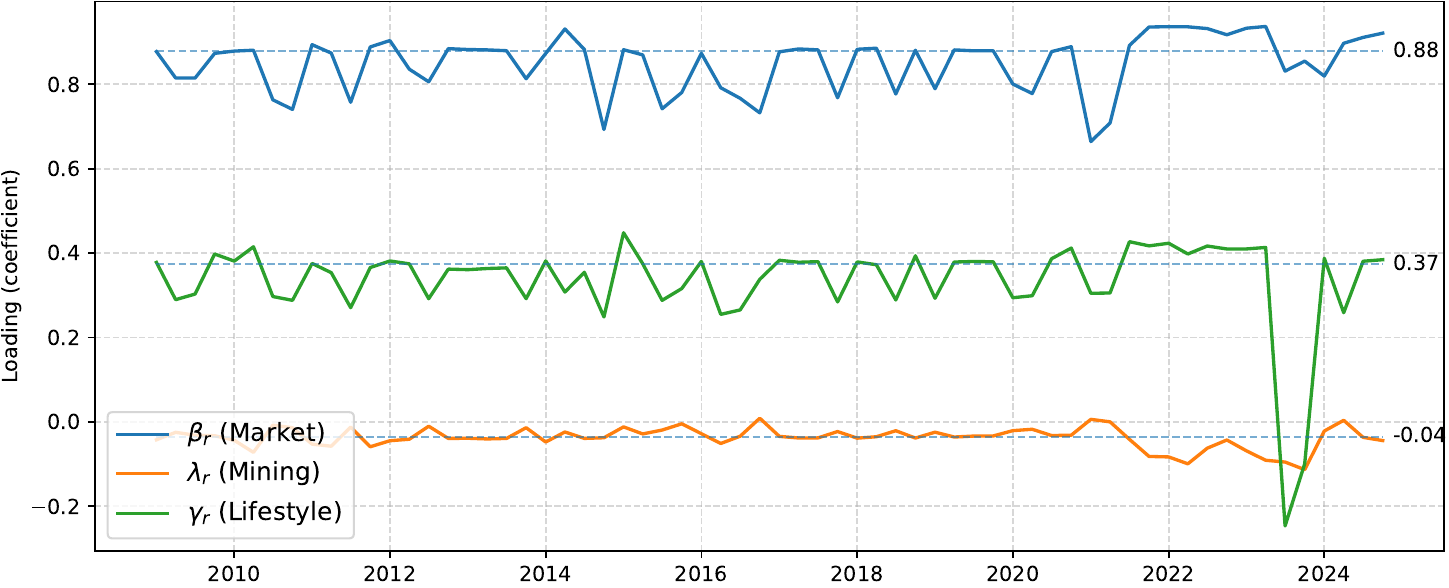}
  \end{subfigure}
  \begin{subfigure}[t]{0.49\linewidth}
    \caption{Rest of QLD}
    \vspace{0.25ex}
    \includegraphics[width=\linewidth]{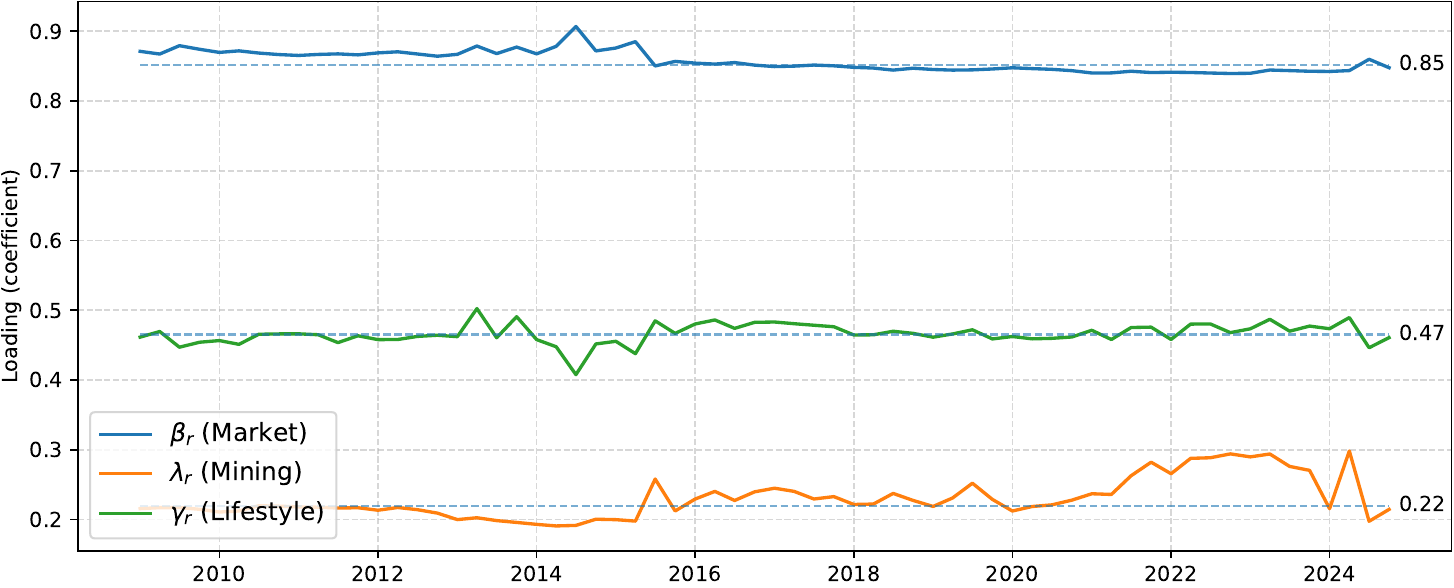}
  \end{subfigure}

  \medskip

  \begin{subfigure}[t]{0.49\linewidth}
    \caption{Greater Adelaide}
    \vspace{0.25ex}
    \includegraphics[width=\linewidth]{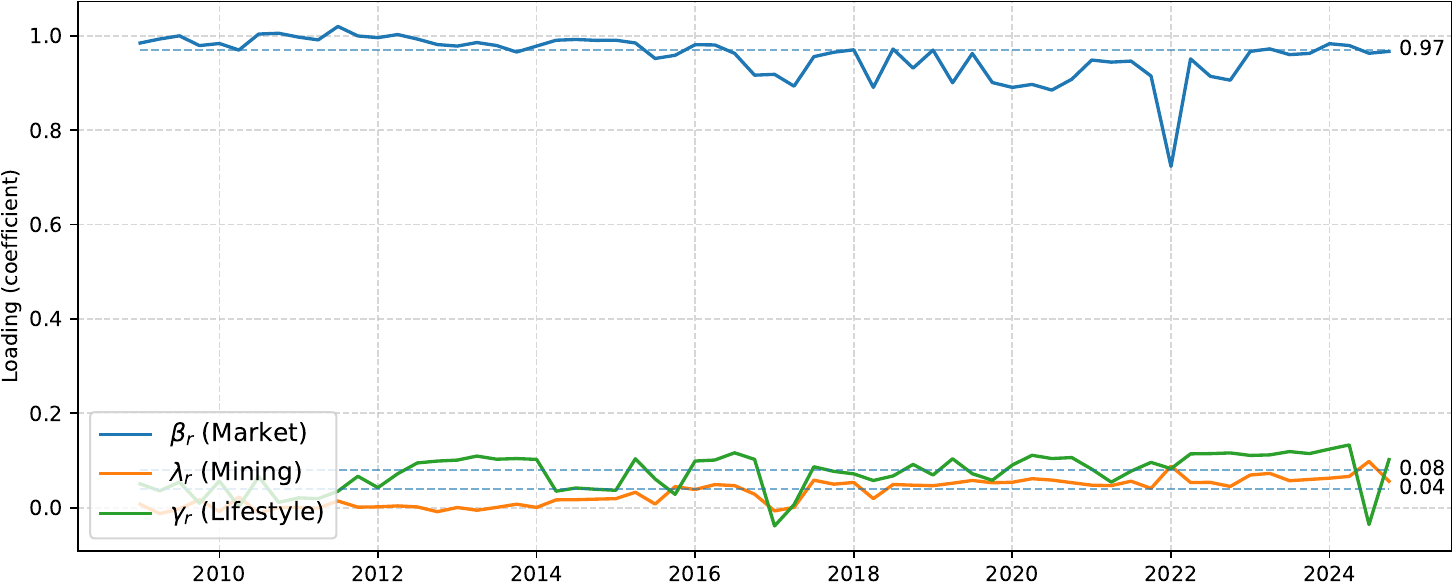}
  \end{subfigure}
  \begin{subfigure}[t]{0.49\linewidth}
    \caption{Rest of WA}
    \vspace{0.25ex}
    \includegraphics[width=\linewidth]{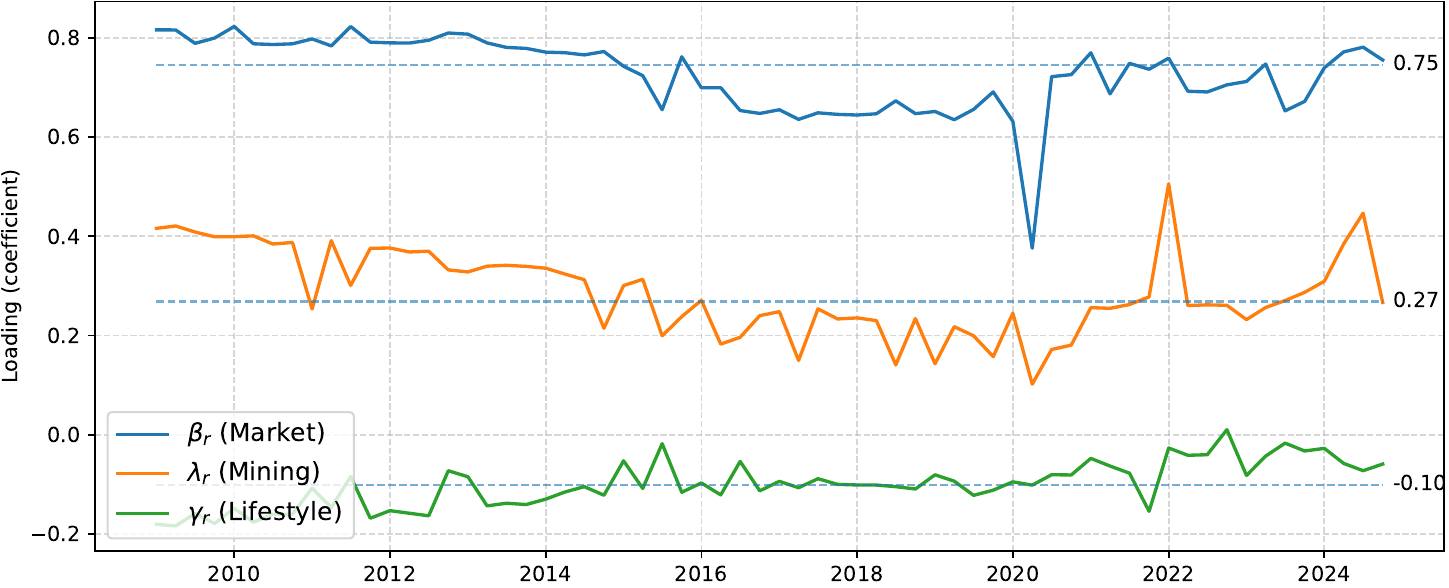}
  \end{subfigure}

  \caption{Expanding-window ARIMAX with seasonal MA(1) (\(s=12\)) factor loadings for major cities. All windows start in January 1995; the endpoint advances from January 2008 to December 2024 in three-month steps. Factor loadings are colored as: Market (\(\beta_r\)): blue; Mining (\(\lambda_r\)): orange; Lifestyle (\(\gamma_r\)): green. }
  \label{fig:expand_panel_cities_topcaps2}
\end{figure}

To test the assumption of our scenario framework that the factor loadings \((\beta_r, \lambda_r, \gamma_r)\) remain stable over time, we estimate the model of Section~\ref{sec:model} on a sequence of expanding windows, each starting in January 1995 and ending at a point we select for testing. We select endpoints between January 2008 and December 2024, advancing in three-month steps. We hold the ARIMAX orders fixed at the values selected in Section~\ref{sec:model_selection} (Table~\ref{app:orders}): these are specifications in levels with seasonal MA(1) at \(s=12\). Figures~\ref{fig:expand_panel_cities_topcaps}–\ref{fig:expand_panel_cities_topcaps2} plot the resulting coefficient paths \((\beta_r,\lambda_r,\gamma_r)\) against the window endpoint.

The loadings exhibit stability over time, albeit with some temporary disturbances or spikiness. For the major cities, the market sensitivity \(\beta_r\) typically varies within a range of \(\pm 0.05\)–\(0.10\) around its median, and the spread loadings \((\lambda_r,\gamma_r)\) show similarly bounded variation consistent with our interpretation of the spreads as stationary processes. Importantly, there is no discernible trend in any of the loadings\footnote{For Brisbane, we interpret the downward excursion for the most recent loadings as a temporary glitch.}, which would signal decreased utility of the factor structure for future expectations. 

To guard against the influence of transient noise in any single window in our conclusions, we calculate each coefficient by its \emph{median} across window endpoints from 2014 to 2024 (we select 2014, as a shorter sample may induce estimation instability without yielding further insights). These median loadings are reported in Table~\ref{table:factor_coefs}. Melbourne stands out with the highest Market beta (\(\beta_r>1\)), Brisbane has an elevated value, while Sydney  and the ACT are close to one and regional areas show dampening (\(\beta_r<1\)) relative to the national market. The Mining and Lifestyle loadings similarly align with qualitative regional narratives (e.g., Perth's strong Mining exposure, Melbourne's negative Lifestyle loading). As a robustness check, Appendix Table~\ref{tab:ci} reports full-sample ARIMAX estimates and 95\% confidence intervals for the factor loadings; these confirm that Melbourne’s Market beta is clearly above one in the statistical sense, whereas the Market betas for Sydney, Brisbane and the ACT are relatively close to one.


\subsection{Examples of price indexes, their decompositions, and relative valuation}
\label{sec:examples_indexes}

\begin{figure}[H]
  \centering
  \includegraphics[width=0.95\linewidth]{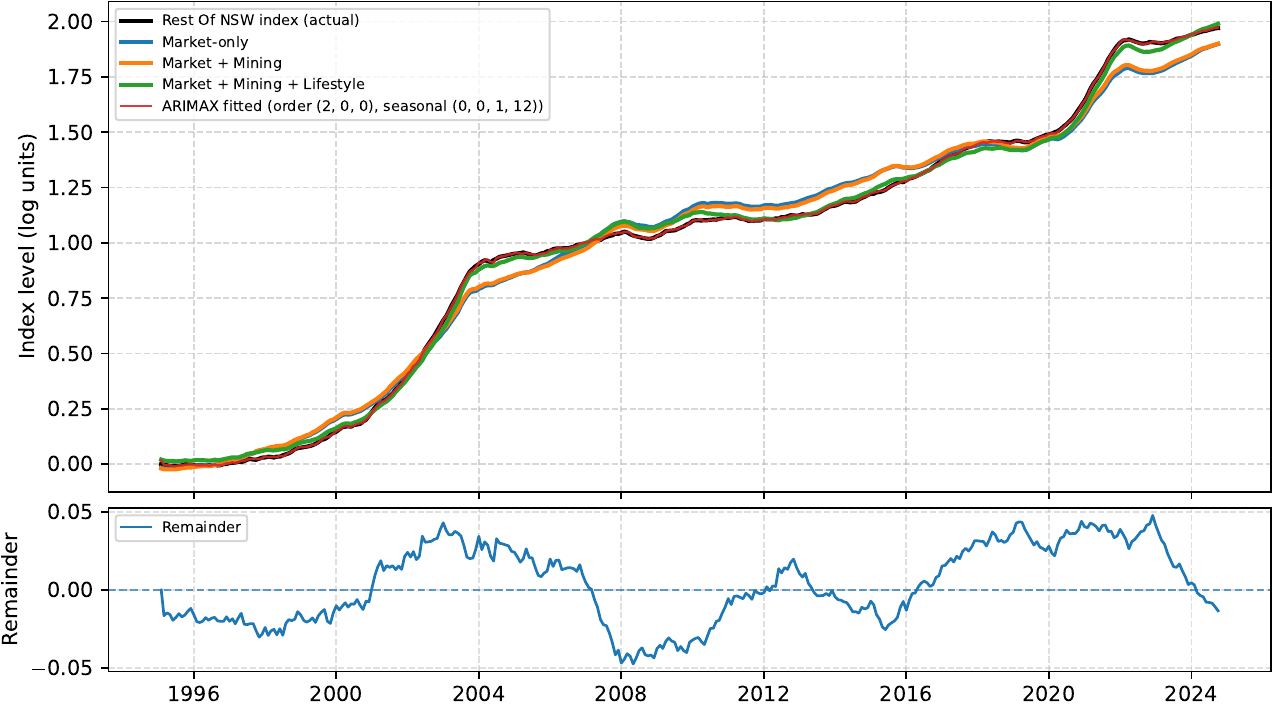}
  \caption{Rest of NSW: Factor-based decomposition of the monthly \emph{log} house price index. 
  Top: observed index, cumulative factor approximations (Market; Market+Mining; Market+Mining+Lifestyle), and the ARIMAX fitted mean under $(2,0,0)$ and $(0,0,1)_{12}$. 
  Bottom: remainder (fitted mean minus three-factor approximation).}
    \caption*{\footnotesize  \textit{Notes}: Positive (negative) factor contributions indicate the market is 
cyclically elevated (depressed) relative to its Market factor baseline.  }
  \label{fig:factor_series_rest_nsw}
\end{figure}

\begin{figure}[H]
  \centering
  \includegraphics[width=0.95\linewidth]{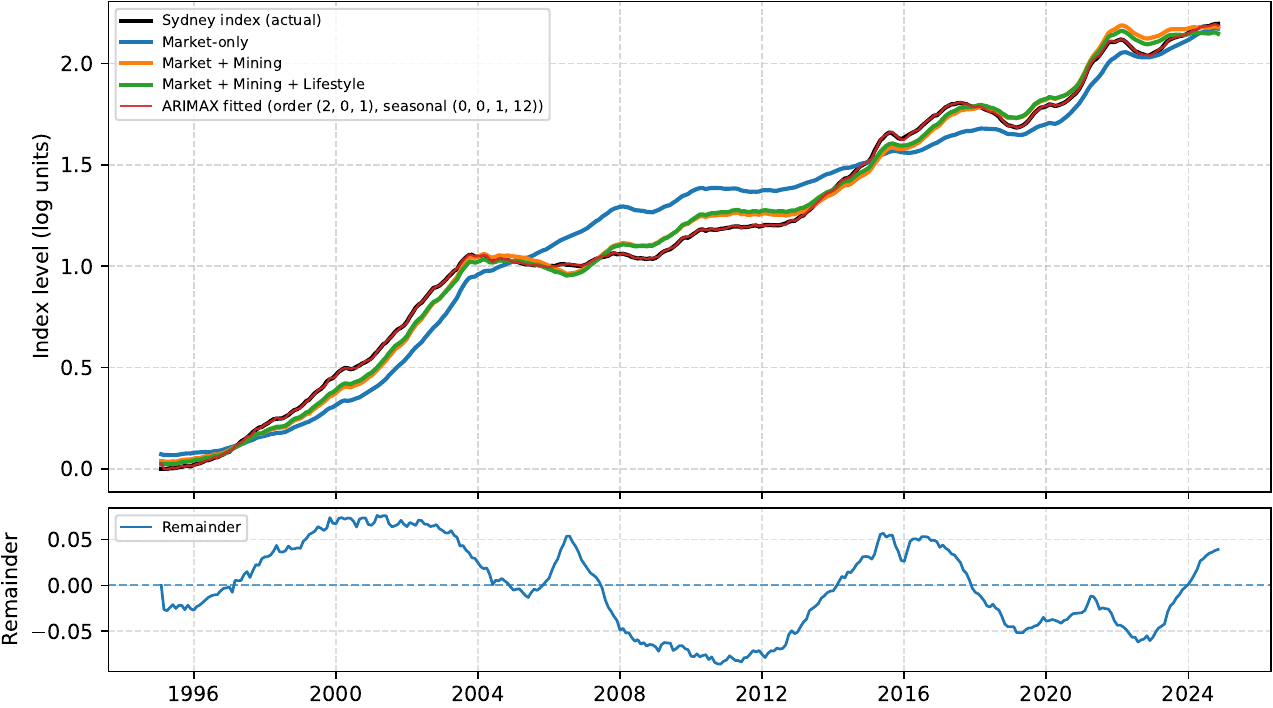}
  \caption{Greater Sydney: Factor-based decomposition of the monthly \emph{log} house price index. 
  Top: observed index (black), cumulative factor approximations (Market; Market+Mining; Market+Mining+Lifestyle), and the ARIMAX fitted mean for $(2,0,1)$ with $(0,0,1)_{12}$. 
  Bottom: remainder (fitted mean minus three-factor approximation).}
  \label{fig:factor_series_sydney}
\end{figure}

\begin{figure}[H]
  \centering
  \includegraphics[width=0.95\linewidth]{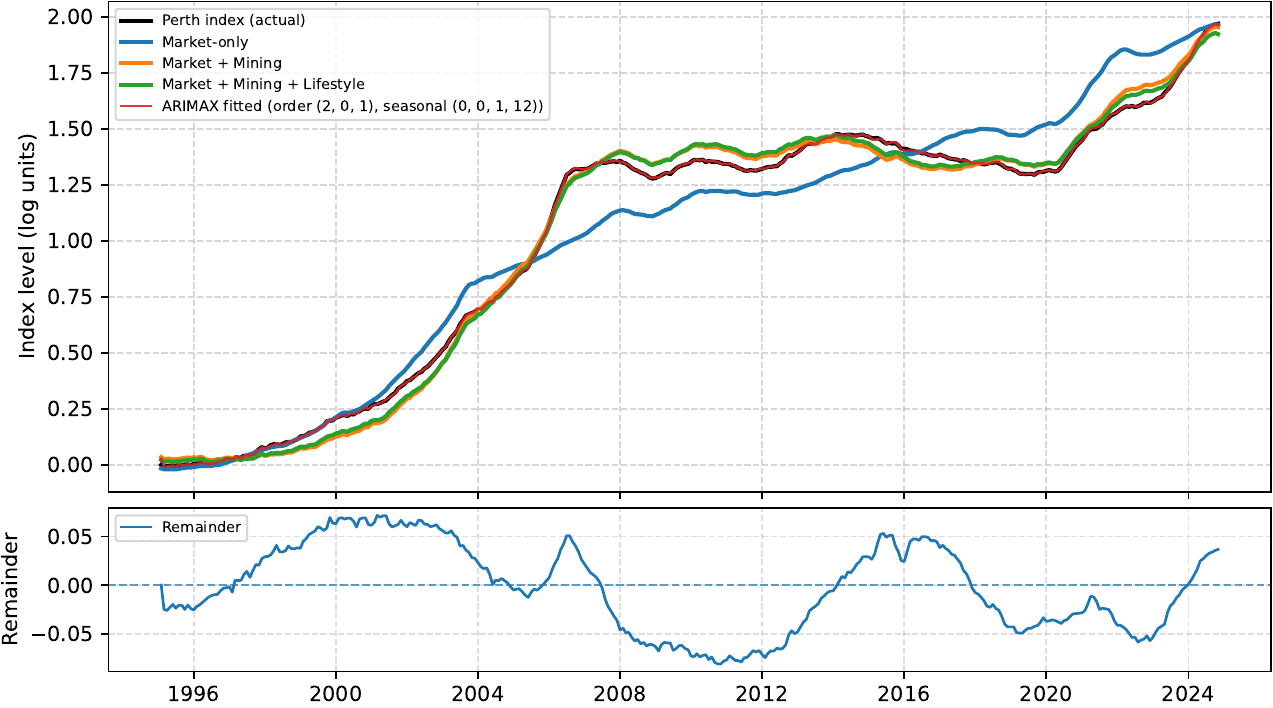}
  \caption{Greater Perth: Factor-based decomposition of the monthly \emph{log} house price index. 
  Top: observed index, cumulative factor approximations (Market; Market+Mining; Market+Mining+Lifestyle), and the ARIMAX fitted mean for $(2,0,1)$ with $(0,0,1)_{12}$. 
  Bottom: remainder (fitted mean minus three-factor approximation).}
  \label{fig:factor_series_perth}
\end{figure}

It is instructive to examine the price index behavior of a few individual price indexes.
Figures~\ref{fig:factor_series_rest_nsw}–\ref{fig:factor_series_perth} illustrate the factor decomposition for three representative markets: the Rest of NSW, with ARIMAX orders $(2,0,0)$, and the major city areas Greater Sydney and Greater Perth, each with ARIMAX orders $(2,0,1)$. Each is estimated with seasonal orders $(0,0,1)_{12}$. The top panel of each figure overlays the observed monthly log house price index with cumulative factor approximations added sequentially (Market; Market+Mining; Market+Mining+Lifestyle) and the ARIMAX fitted mean. As factors are added, the approximation tightens; with all three factors the fitted mean is nearly indistinguishable from the observed index at the plotting scale. 

The bottom panel reports the remainder $\epsilon_{r,t}$, defined as the ARIMAX fitted mean minus the three-factor approximation. Across all three markets this remainder exhibits no visual drift, consistent with the factors capturing the low-frequency co-movement, and shows a gentle undulating pattern consistent with an ARMA component plus seasonal MA(1) at 12 months, supporting the $(2,0,1)$ with $(0,0,1)_{12}$ specification. In addition, the Ljung--Box $p$-values in Table~\ref{table:lifestyle_inclusion} are generally above 0.05 for the selected specifications, indicating no evidence of remaining autocorrelation up to 12 lags. 

\paragraph{Rest of NSW: Lifestyle-driven cycles.}
The Rest of NSW (Figure~\ref{fig:factor_series_rest_nsw}) is commonly characterized as a lifestyle market with limited direct exposure to mining. This is reflected in the decomposition of its log price index: despite the presence of mining regions within NSW (e.g. Singleton), the Mining factor makes little difference to the Market-only approximation (in fact, the loading is slightly negative, indicating a Sydney influence), whereas adding the Lifestyle factor (green series) explains a large share of the overall growth pattern. As in other lifestyle regions, this imparts a more cyclical profile: a sharp run-up into the mid-2000s peak (circa 2004), a prolonged plateau through roughly 2012, and renewed growth that accelerates during the COVID-19 period beginning in 2020.


\paragraph{Sydney: Negative mining exposure and the terms-of-trade boom.}
The Sydney log price index (Figure~\ref{fig:factor_series_sydney}) is not strongly influenced by Lifestyle; instead, adding the Mining factor (orange series) materially improves the fit compared to the market-only approximation (blue), reflecting Sydney's \emph{negative} Mining loading ($\lambda_{\text{Sydney}}=-0.43$). This pattern is apparent in \citet{sijp_francke_ree2024}, and remains somewhat unexplained. One context is Australia’s terms of trade (ToT). Using national quarterly data, \citet{tumbarello_wang2010} find that positive ToT shocks raise Australian real house prices in the long run, while \citet{downes_et_al_rba_report2014} provide a detailed analysis of the mining boom. 

In our decomposition, the Australia-wide effect of this ToT/mining-boom impulse is absorbed into the Market factor, and the Mining spread models local variation around this common influence: Figure~\ref{fig:factor_series_sydney} shows Sydney underperforming its Market-factor baseline through the mid-2000s boom phase, consistent with $\lambda_{\text{Sydney}}<0$. This growth during 2004--2006 is quite flat, and could reflect a reduced sensitivity of Sydney to the mining boom, or a net negative effect of the resource cycle on Sydney price growth via channels such as internal migration and local demand and supply conditions. Australian evidence suggests that house prices have become increasingly sensitive to mortgage rates, with the largest marginal effects in Sydney and Melbourne \citep{otto2007}. In addition, Sydney and Melbourne are more exposed than the other capitals to foreign real-estate investment, and comparative work on large cities emphasises the growing role of global investors and financial conditions in driving price dynamics in major urban housing markets.\footnote{See, for example, the discussion of foreign investors and global financial conditions in large cities in \citet[][ch.~14]{Nijskens2019HotProperty}.} Insofar as the mining boom pushed the Australian dollar to historically high levels and was associated with higher interest rates, Sydney housing would have become more expensive to foreign-currency buyers precisely when domestic financing conditions were tightening. This combination could plausibly have dampened Sydney’s response to the mining boom, with a similar but weaker effect in Melbourne.


\paragraph{Perth: Positive mining exposure.}
A pattern converse to Sydney appears for Perth (Figure~\ref{fig:factor_series_perth}). The mining boom lifts prices relative to Perth's Market factor (consistent with a strong positive loading, $\lambda_{\text{Perth}} = 0.60$), with a subsequent decline after the boom's end in 2013 and a reversal beginning around 2020 marked by a sharp increase thereafter. Unlike the Lifestyle factor, the Mining factor explains a substantial share of the deviation of Perth's index from the Market path, reflecting the mining sector's influence on employment, population flows, incomes, and the local economy.


The three cities shown in this section do not exhibit strong changes in their reliance on the three factors over time: the magnitude of the remainder is relatively constant in time, suggesting that city-specific behavior has not increased or decreased over time. Perth and Sydney have similar underlying growth, we will see in Section~\ref{sec:regional_amplification} that both roughly track the market factor with a $\beta_r \approx 1$, with Perth somewhat lower than Sydney (this is also visible in Figures~\ref{fig:factor_series_sydney},\ref{fig:factor_series_perth}). Therefore, a large part of the differences in price behavior between these two cities can be explained by the opposite-phase of the Mining factor, strong in both cities.\footnote{The price indexes in this paper concern growth relative to a reference time, not actual price levels such as a median city price. Naturally, the cities differ considerably in median house price, with Sydney's median well exceeding Perth's.}

The decomposition allows an attribution of departures from the longer-term growth path to economically meaningful processes, providing a framework for assessing whether a local market is ``overvalued'' or ``undervalued'' relative to its typical relationship with the national Market factor. For instance, in 2024 the Rest of NSW price is higher than would be expected based on the Market factor alone (Figure~\ref{fig:factor_series_rest_nsw}). This departure is attributable to the Lifestyle factor, which remains elevated during this period. Compared to historical patterns, this positive deviation indicates higher than usual price growth driven by the Lifestyle component, consistent with the documented increase in internal migration away from inner cities post-COVID, enhanced by emerging remote work opportunities \citep{yanotti2023}.

Conversely, Sydney house price growth was subdued during the 2003--2012 mining boom period, as reflected by its negative loading on the Mining factor ($\lambda_{\text{Sydney}} = -0.43$). This rendered Sydney historically undervalued compared to its Market factor trajectory during this period (Figure~\ref{fig:factor_series_sydney}), creating potential investment opportunities---an inverted image of Perth's mining-driven overvaluation during the same years ($\lambda_{\text{Perth}} = 0.60$). This illustrates how the factor decomposition not only identifies the sources of price movements but also provides a quantitative basis for relative valuation assessments across markets and time periods, distinguishing cyclical (mean-reverting) effects from fundamental Market-driven growth.

\subsection{Regional amplification and long-run growth differentials}
\label{sec:regional_amplification}

We now examine price growth, where a typical house in region $r$ valued at some price $p$ at some time $t_1$ appreciates to $f_rp$ at some later time $t_2$ (following the price index of that region), with growth $f_r = e^{\mu({t_2})-\mu({t_1})}$, and compare it to growth in the national index over the same period, $f_M = e^{U({t_2})-U({t_1})}$. Of particular interest is the market-only approximation $\mu_r \approx \beta_rU$.
The market loadings \(\beta_r\) in Table~\ref{table:factor_coefs} reveal substantial and persistent growth differentials across Australian housing markets. A region with \(\beta_r > 1\) amplifies national market movements, while \(\beta_r < 1\) dampens them. Over the 30-year sample period, the national market index exhibits a doubling time of approximately 10 years, corresponding to a nominal compound annual growth rate (CAGR) of 7.2\%, consistent with one Reserve Bank estimates of roughly 7.25\% per annum over the three decades to 2015 \citep{kohler_vandermerwe_2015}. The regional \(\beta_r\) values transform this national trend into markedly different long-run trajectories.

\paragraph{The amplification hierarchy.}
Melbourne exhibits the strongest amplification (\(\beta_r = 1.22\)), implying that a national doubling translates into a 2.33-fold increase (column \(f_r|f_M=2\) in Table~\ref{table:factor_coefs}) and a doubling time of just 8.2 years on historical growth (CAGR: 8.8\%). Brisbane follows with \(\beta_r = 1.10\), yielding a 2.14-fold increase under national doubling and a 9-year doubling time (CAGR: 7.9\%). Sydney tracks at a similar pace to the national market by its market projection (\(\beta_r = 1.01\), doubling time 9.9 years, CAGR: 7.2\%), while the remaining state capitals, Adelaide, Perth, Hobart, show modest dampening  (\(\beta_r \approx 0.91\)–\(1.02\)). 

The regional areas exhibit substantially weaker sensitivities. Rest of NSW (\(\beta_r = 0.88\)) and Rest of WA (\(\beta_r = 0.70\)) dampen national movements considerably, with doubling times of 11–14 years and CAGRs in the range 4.9–6.2\%. Under a national doubling, these regions increase by only 60–85\%. This cross-sectional pattern, where major cities amplify, regional areas dampen, is economically intuitive: large metropolitan markets face tighter supply constraints, attract stronger migration flows, and exhibit greater sensitivity to credit and sentiment cycles than do dispersed regional markets. 

\paragraph{Which elements are stable?}
Naturally, it remains unknown whether the strong growth in the market factor path itself will persist into the coming decades. 
On internationally used price-to-income metrics, Australia’s major cities (especially Sydney) sit among the least affordable in the world \citep[e.g.][]{demographia2025,anzcorelogic2024}, which plausibly points to slower national growth ahead (i.e., longer doubling times) than in recent decades. This future national development does not impact our conclusions, as our stability claim is narrower: the \emph{loading coefficients} (e.g., $\beta_r$), and the way Mining and Lifestyle enter as stationary spreads around the market path, appear stable. Thus, when we map a hypothetical national change $f_M$ into $f_M^{\beta_r}$ for region $r$, we intend a \emph{relative scaling rule} conditional on whatever national path one assumes. In contrast to the loadings, the uncertainty band widths related to the Mining, Lifestyle and idiosyncratic factors (Eq.~\ref{eq:uncertainty}, and discussed below) associated with a national doubling \emph{do} depend on the Market factor as they depend on the doubling time: a rapidly changing market will have narrower bands for the bounded factors (as doubling is achieved sooner, so that spread uncertainty has had less time to build).

\paragraph{Uncertainty decomposition.}
The market-only projection \(f_r = f_M^{\beta_r}\) provides a baseline scenario, but regional outcomes deviate due to the Mining and Lifestyle spreads and the idiosyncratic remainder \(\epsilon_r\) (Eq.~\ref{eq:uncertainty}). Column x95 in Table~\ref{table:factor_coefs} reports the 95\% multiplicative band arising from the two factor spreads alone at the 10-year horizon. For most cities this band is modest: 1.03–1.26, indicating that the market component dominates the long-run trajectory. Perth is the notable exception (x95 = 1.38), reflecting its large positive exposure to the Mining spread (\(\lambda_{\text{Perth}} = 0.60\)). 

Adding the idiosyncratic component at the 10 year horizon (column x95 total) widens the bands slightly. Brisbane (1.14), Adelaide (1.10), and ACT (1.10) exhibit the tightest total bands among major cities, while Melbourne (1.25), Sydney (1.27), and Hobart (1.30) show moderate widening. Perth (1.39) and Darwin (1.39) are outliers with substantially wider uncertainty, driven by Mining exposure in Perth's case and idiosyncratic volatility in Darwin's.
Here, we chose the 10-year horizon for the uncertainties to match the observed 10 year historical national house price doubling time so as to be able to compare the columns in the table.
\footnote{Here, ceteris paribus, the behavior of the market $U$ does control the overall uncertainty band widths (Eq.~\ref{eq:uncertainty}), e.g. a rapidly rising national market will have narrower uncertainty bands for the bounded factors as Market factor doubling is achieved more quickly, allowing less time for the remaining factors to evolve.}

\begin{figure}[H]
    \begin{center}
    \scalebox{1}{\includegraphics[width=0.95\linewidth]{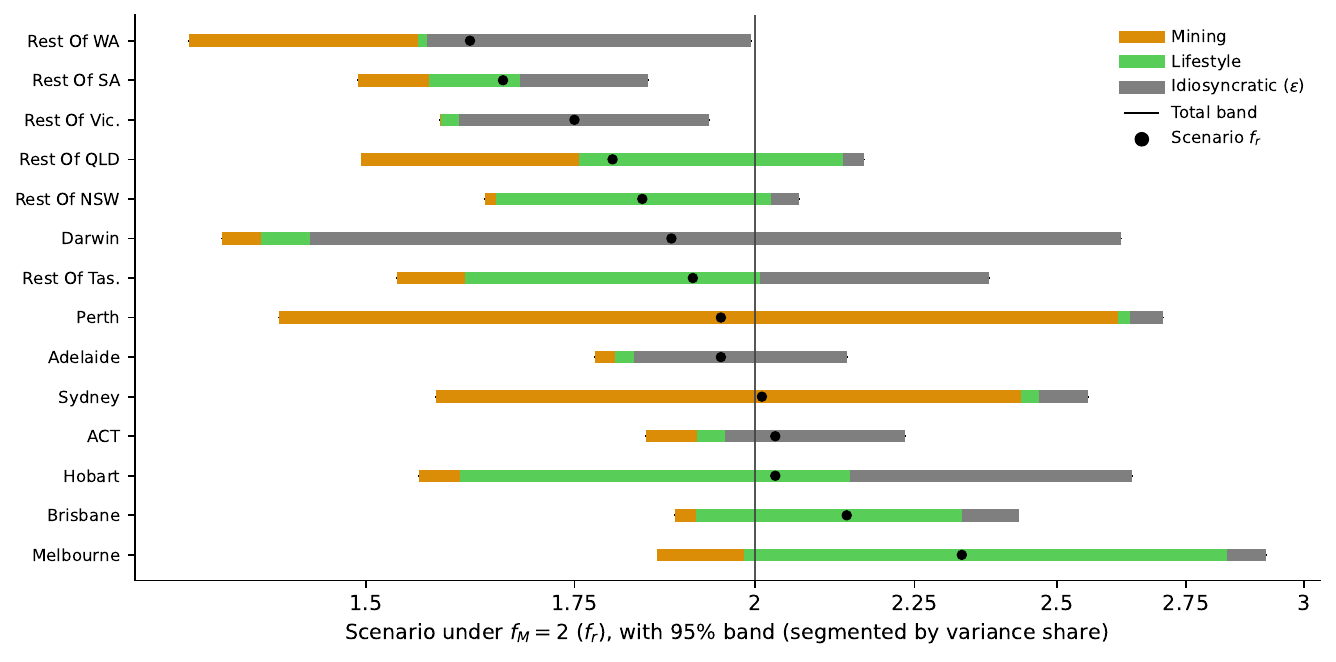}}
    \end{center}
\caption{Scenario mapping under a national doubling ($f_M=2$), with the 95\% uncertainty band decomposed by source. For each region, the horizontal bar spans $[\,f_r/x95_{\text{total}},\,f_r\times x95_{\text{total}}\,]$ where $f_r=2^{\beta_r}$. Colored segments indicate variance shares (log space) from Mining (brown), Lifestyle (green), and idiosyncratic $\epsilon$ (grey);  the dot marks $f_r$. Regions are sorted by $\beta_r$; the vertical line marks $f_M=2$. }

    \label{fig:scenario_bands}
\end{figure}

\paragraph{Interpreting the factor loadings.}
Figure~\ref{fig:scenario_bands} visualizes the scenario under a national doubling (\(f_M=2\)), with the 95\% band decomposed by source (based on a 10 year horizon corresponding to the 30-year historical doubling time). Perth's large Mining contribution (brown segment) dominates its uncertainty, arising from its positive loading (\(\lambda_r = 0.60\)) on the Perth–Sydney spread, and its overall variability is large compared to the other cities, emphasizing the more volatile nature of Perth's housing market. This spread captures the resource-sector cycle; when mining booms, the spread rises and Perth benefits directly through this positive exposure. Sydney, by contrast, has a large \emph{negative} loading on Mining (\(\lambda_r = -0.43\)), meaning it moves counter-cyclically to the resource boom: as the spread rises (Perth outperforms), Sydney underperforms relative to the market factor. This opposing-phase behavior explains why both cities show strong Mining factor contributions to their variance in Figure~\ref{fig:scenario_bands}, despite being on opposite ends of the spread.   
Melbourne exhibits the strongest fundamental growth (\(\beta_r=1.22\)) and a significant negative loading on the Lifestyle factor (\(\gamma_r = -0.70\)), making Lifestyle the dominant non-market source of variation. This negative exposure suggests that Melbourne tends to move counter-cyclically with respect to amenity-driven demand shifts. As also described in Section~\ref{sec:pca}, this is consistent with longstanding interstate migration patterns from the southern states to Queensland (which loads highly on the Lifestyle factor; see below). Several demand-side factors provide a plausible explanation: relatively high housing costs, a colder and more variable climate, which for some households may be less attractive than warmer coastal regions, and the increased feasibility of remote work, all of which may encourage some households to relocate to more affordable coastal areas \citep[][]{yanotti2023}. Melbourne’s relative underperformance in recent years also coincides with a period of extended COVID-19 restrictions and subsequent changes to property taxation, both of which may have influenced household location and tenure decisions.

Brisbane combines high fundamental growth (\(\beta_r=1.10\)) with tight uncertainty bands (x95 total = 1.14). Its positive Lifestyle loading (\(\gamma_r = 0.39\)) puts this city with a warmer climate in contrast to Melbourne, and is consistent with its position as a lifestyle-migration destination. Mining plays almost no role (\(\lambda_r = 0.06\)). Important other recipients of the lifestyle migration push out of the southern cities are the regional areas Rest of QLD (\(\gamma_r = 0.47\)), Rest of NSW (\(\gamma_r = 0.38\)), and Rest of Tas (\(\gamma_r = 0.54\)). They are dominated by positive Lifestyle loadings, as expected for amenity-rich coastal and hinterland markets, e.g. Bateman's Bay in NSW and Sunshine Coast and Surfers Paradise in QLD. They all show strong positive exposures, while their Mining loadings remain modest, except in Rest of QLD, where mining is also a significant factor alongside lifestyle, as is expected from the large mining operations for instance in Central Queensland. The idiosyncratic component is relatively small for Rest of QLD and Rest of NSW, indicating that most of the cyclical variability there is captured by the Mining and Lifestyle factors.
In contrast to most of the cities, Adelaide's volatility largely arises processes other than the Mining and Lifestyle factors of our model, suggesting a more idiosyncratic character for this city.

On current trends, the stable $\beta_r$ hierarchy implies widening long-run level differences: if the national factor continues to rise, higher-$\beta_r$ cities will outpace lower-$\beta_r$ regions. The same scaling works in reverse in a downturn or stagnation: for any national path $f_M<1$, high-$\beta_r$ markets (e.g., Melbourne) contract more, while lower-$\beta_r$ regional areas are less sensitive than the capitals and the national aggregate.


\begin{table}[ht!]
	\centering
    \small

\begin{tabular}{lrrrrrrr}
\toprule
Region & $\beta_r$ & $\lambda_r$ & $\gamma_r$ & $f_r|f_M=2$ & Doubling time & x95 & x95 total \\
\midrule
Melbourne & 1.22 & -0.16 & -0.70 & 2.33 & 8.16 & 1.24 & 1.25 \\
Brisbane & 1.10 & 0.06 & 0.39 & 2.14 & 9.05 & 1.12 & 1.14 \\
Hobart & 1.02 & 0.12 & 0.68 & 2.03 & 9.76 & 1.23 & 1.30 \\
ACT & 1.02 & -0.08 & 0.11 & 2.03 & 9.76 & 1.05 & 1.10 \\
Sydney & 1.01 & -0.43 & -0.14 & 2.01 & 9.86 & 1.26 & 1.27 \\
Adelaide & 0.96 & 0.05 & 0.09 & 1.95 & 10.37 & 1.04 & 1.10 \\
Perth & 0.96 & 0.60 & -0.13 & 1.95 & 10.37 & 1.38 & 1.39 \\
Rest Of Tas. & 0.93 & 0.14 & 0.54 & 1.91 & 10.71 & 1.19 & 1.24 \\
Darwin & 0.91 & 0.13 & -0.27 & 1.88 & 10.94 & 1.11 & 1.39 \\
Rest Of NSW & 0.88 & -0.04 & 0.38 & 1.84 & 11.31 & 1.12 & 1.12 \\
Rest Of QLD & 0.85 & 0.23 & 0.47 & 1.80 & 11.71 & 1.20 & 1.20 \\
Rest Of VIC & 0.81 & 0.01 & 0.09 & 1.75 & 12.29 & 1.03 & 1.10 \\
Rest Of SA & 0.73 & 0.10 & 0.21 & 1.66 & 13.64 & 1.08 & 1.11 \\
Rest Of WA & 0.70 & 0.25 & -0.09 & 1.62 & 14.22 & 1.14 & 1.23 \\
\bottomrule
\end{tabular}
 \caption{Regional market sensitivity and 10-year uncertainty.}
\begin{tablenotes}[flushleft]\footnotesize
\item $f_r|_{f_M=2}=2^{\beta_r}$; Doubling time $T_r=T_M/\beta_r$. 
x95 are 10-year multiplicative 95\% bands; “total” adds idiosyncratic $\epsilon$. 
Loadings are medians across expanding-window ARIMAX fits (orders in Appendix Table~\ref{app:orders}).
\end{tablenotes}
    \label{table:factor_coefs}
\end{table}

%

\section{Conclusion}
\label{sec:conclusion}

We studied long-run growth differences across Australian house markets (1995–2024, houses only) using a transparent three-factor model in levels: a Market factor that absorbs the main trend common to all indexes up to a coefficient, plus two more cyclical spreads (Mining and Lifestyle) that capture mean-reverting departures from the national path. These interpretable near-orthogonal factors proxy the first three principal components of the granular local price indexes, explaining most of the variance in that dataset.
This near-independence allows straightforward ARIMAX estimation and interpretation. 
We have shown that the market factor, i.e. the city-specific scaling of the national price index, explains significant parts of the price developments in the cities. Conversely, volatility beyond the market factor is largely explained by the Mining and Lifestyle factor for the most important cities and regions, with Adelaide and ACT among the exceptions, where departures arise largely from idiosyncratic movement specific to each city.

Regional market sensitivities (\(\beta_r\)) are stable in expanding windows, yielding a persistent cross-sectional ranking: large capitals track or amplify national movements, while regional areas have a damped relationship to the national market.  The Mining and Lifestyle spreads behave as stationary factors: they widen forecast fans but do not overturn the long-run ordering implied by the market sensitivity.  
Scenario mapping is therefore simple and transparent: for any assumed national change by a factor \(f_M\), the market-only regional baseline is \(f_M^{\beta_r}\), with spread and idiosyncratic variances adding a band.
Ten-year bands are generally moderate once scaled by loadings, with wider ranges only where exposures are large (e.g., Perth to Mining) or idiosyncratic volatility is high. Decomposing the bands clarifies whether Mining, Lifestyle, or the remainder drives regional risk around the market-only path. Specifically, Perth shows high volatility outside the market factor, dominated by the Mining factor. Perhaps more surprisingly, this is also the case for Sydney, through its opposing price behavior.


Beyond characterizing long-run growth differentials, the factor model provides a practical framework for assessing relative valuation across markets and time, as we can identify when regional markets are historically elevated or depressed relative to their long-run paths in the context of these large-scale geographic price movements. 
For example, in 2024, the Rest of NSW exhibits prices higher than its Market factor trajectory would suggest, attributable to an elevated Lifestyle factor consistent with sustained post-COVID migration to amenity-rich coastal areas. 
The factor loadings provide objective metrics for assessing whether current local price levels reflect long-term growth in line with the national market, or variations of a more cyclical nature and of geographic specificity.
This complements traditional valuation metrics (e.g., price-to-income ratios). 
The decomposition does not assess whether the Market factor itself is overvalued in an absolute sense (e.g., relative to income fundamentals or long-run sustainable levels).
While absolute affordability measures indicate whether national or regional housing is expensive relative to fundamentals, the factor decomposition reveals whether a market is expensive or cheap \emph{relative to its own historical relationship with the national Market factor}. 
Naturally, the national market itself undergoes cycles, but these are experienced by all regions to varying degrees according to their $\beta_r$. Unlike the Mining and Lifestyle cycles, whose effects can partially offset across regions with opposing loadings, the national Market component is systematic and cannot be diversified away by geographic spreading.

With the cyclical factors removed, Melbourne leads the major cities with significantly higher amplification of the market factor, followed by Brisbane.  Sydney, Adelaide and Perth are more on-par with the national pace (upon removal of the cyclical factors), whereas the regions lag. 
Given that the eight capital cities collectively account for roughly two-thirds of national house sales in our sample, the national repeat-sales index is significantly influenced by the paths of Sydney, Melbourne and, to a lesser extent, Brisbane (Table~\ref{table:counts}).  International evidence suggests that housing markets in large cities are characterized by limited supply (with limited immediate scope for infill developments, particularly for houses), building restrictions, slow supply response times and geographic constraints. The resulting low elasticity allows steep house price increases driven by high demand from the pull to big cities generated by employment and education opportunities and social, cultural and recreational amenities. Migration and decreasing household size add to local population pressures \citep{kohler_vandermerwe_2015}. With housing stock and wealth concentrated in these cities, price movements tend to ripple out to surrounding regions and help lead the national cycle \citep[e.g.][]{Nijskens2019HotProperty,otto2007}. The well-documented increasing popularity and rising house prices of big cities throughout the world \citep{Nijskens2019HotProperty} is reflected in the higher betas of the Australian cities found in our study. Complementary to this, the sustained increases in the national price index of the last 30 years, with the most recent growth period starting around 2012 reflect the big cities combined. This arises from a long-term global urbanization process, where work opportunities are increasingly concentrated in the larger cities.
The regions lag behind with an increasing price gap, as reflected in their low betas.

\citet{otto2007} finds that real mortgage rates are an important determinant of Australian city-level house price growth and that cities differ in their sensitivity.
\citet{he_lacava_rba2020} find that this  sensitivity is greater in Australian cities where land is more expensive, where incomes are relatively high, households are more indebted and there are more investors. 
This is consistent with the widely held view that monetary policy can have heterogeneous regional effects. Although such mechanisms provide some background to the differences in $\beta_r$ that we find, a full exploration is beyond the scope of this paper.

Several limitations of our approach suggest natural extensions. We impose a low-dimensional, three-factor model with interpretable proxies for Mining and Lifestyle; while this matches the leading principal components of the granular indexes, additional factors or direct use of the PCA series could capture further geographic nuances. The analysis is based on repeat-sales house indexes, so units (apartments), may exhibit different sensitivities. Finally, relating the estimated factor loadings and time series to fundamentals such as interest rates, incomes and supply, is a natural direction for future work.

  Long-term geographical variations in appreciation have persisted in the Australian housing market. 
  This study has attempted to disentangle the underlying market factor from the more cyclical and geographically distributed processes in these growth patterns.
Naturally, there is no guarantee that the growth differences that characterized the last 30 years will persist in the future. Our contention, however, is that there is no trend-based indication in the house price data as yet that this will be the case, neither in the expected time development of the loadings, nor in the patterns exhibited by the ARMA-modeled city-specific remainder. This suggests that our growth pattern characterizations of the major cities and regions will persist into the next 5--10 years.

\textit{Acknowledgments:} I thank Yiran Yao (Monash Master of Business Analytics) for research assistance, text editing, stimulating discussions, data processing and creating \ref{app:mining_regimes} during her internship at NEOVAL. I also thank the participants of the 2025 Time Series and Forecasting Symposium at the University of Sydney for their helpful feedback on my presentation of this study.

\textit{Disclosure:} No conflicts of interest. Views are those of the author and not necessarily of Neoval Pty Ltd or UTS.

\clearpage
\begingroup
\bibliographystyle{elsarticle-harv}
\setlength\bibsep{0.5pt}
\bibliography{refs}
\endgroup

\clearpage
\appendix
\renewcommand{\thesection}{Appendix~\Alph{section}}

\section{ARIMA orders for the cities and regions}


\begin{table}[ht!]
\centering
\small
\begin{tabular}{lcc}
Region & ARIMA(p,d,q) & Seasonal (P,D,Q)[s] \\
ACT & (1,0,0) & (0,0,1)[12] \\
Adelaide & (2,0,1) & (0,0,1)[12] \\
Brisbane & (2,0,1) & (0,0,1)[12] \\
Darwin & (1,0,2) & (0,0,1)[12] \\
Hobart & (1,0,1) & (0,0,1)[12] \\
Melbourne & (2,0,2) & (0,0,1)[12] \\
Perth & (2,0,1) & (0,0,1)[12] \\
Rest Of NSW & (2,0,0) & (0,0,1)[12] \\
Rest Of QLD & (1,0,2) & (0,0,1)[12] \\
Rest Of SA & (2,0,0) & (0,0,1)[12] \\
Rest Of VIC & (2,0,0) & (0,0,1)[12] \\
Rest Of WA & (2,0,0) & (0,0,1)[12] \\
Sydney & (2,0,1) & (0,0,1)[12] \\
\end{tabular}
\caption{Selected ARIMAX orders for regional house price indexes (three–factor specification). All models are estimated in levels with a constant ($d=0$ and $D=0$) and include exogenous factors $U_t$, $\delta_{PS,t}$, and $\delta_{L,t}$. The seasonal component is fixed at MA(1) with period 12. For each region, non-seasonal orders $(p,q)$ are chosen by minimising AICc over the grid $\{0,1,2\}\times\{0,1,2\}$. Most selections are $(2,0,1)$ or $(2,0,0)$ with MA(1)[12].}
\label{app:orders}
\end{table}

\section{Underlying price indexes}
\label{sec:underlying_indexes}

Our factor analysis requires monthly house price indexes at multiple geographic scales. We construct these indexes using a repeat sales regression, based on \citet{bailey:1963} but with spatially regularized priors. The model is identical to that used and described in \citet{sijp_francke_laplacian}, and we will briefly describe it here for completeness.

\paragraph{Index construction at the SA2 level.}
At the finest SA2 geography (2,052 regions), transaction volumes are often too sparse for reliable unregularized repeat sales estimation, a method that requires sufficient volumes of sales data \citep{malone_ree2020, bogin_doerner_larson_faj2019, contat_larson_ree2024}. We address this using a Bayesian framework that pools information across neighboring regions via graph-based spatial priors. The model is estimated separately for each SA4 area (comprising ~25 SA2 regions on average), allowing indexes to borrow strength from spatially proximate markets while preserving local variation.

The repeat sales regression takes the standard form
\begin{equation}
    \boldsymbol{y} = \boldsymbol{j} \theta  + D^{(t)} \boldsymbol{\mu} + D^{(s,t)} \boldsymbol{\alpha} + \boldsymbol{\varepsilon}, \quad 
    \boldsymbol{\varepsilon} \sim \mathcal{N}(0, \sigma^2 I), \label{eq:stmM_old}
\end{equation}
where $\boldsymbol{y}$ contains log price changes for repeat sales, $\boldsymbol{j}$ is a vector of ones (intercept), $D^{(t)}$ is the time selection matrix (±1 for sale/purchase dates), and $D^{(s,t)}$ is the space-time selection matrix. The vector $\boldsymbol{\mu}$ represents the common time trend, while $\boldsymbol{\alpha} = \text{vec}(\tilde{\alpha})$ captures local deviations from this trend across SA2 regions.

We impose regularizing priors that penalize abrupt changes in time and space. Following \citet{goetzmann:1992}, the common trend follows a random walk prior, $\boldsymbol{\mu} \sim \mathcal{N}(0,\sigma_{\mu}^2 L^{(t)+})$, where $L^{(t)}$ is the temporal graph Laplacian (differences between consecutive months). Spatial structure is captured via the adjacency graph of SA2 regions, encoded in the spatial Laplacian $L^{(s)} = \Delta - W$, where $\Delta$ is the degree matrix (diagonal with node degrees) and $W$ is the adjacency matrix (unit weights for neighboring regions). The local deviations $\boldsymbol{\alpha}$ receive a spatio-temporal prior $\boldsymbol{\alpha} \sim \mathcal{N}(0,\sigma_{\alpha}^2 L^{(s,t)+})$ with $L^{(s,t)} = L^{(s)} \otimes L^{(t)}$.\footnote{The superscript $+$ denotes the Moore-Penrose generalized inverse, coinciding with the standard inverse for full-rank matrices.}

In a penalized regression framework, this Bayesian specification is equivalent to minimizing
\begin{equation}    
    \|\boldsymbol{y} - \boldsymbol{j} \theta - D^{(t)} \boldsymbol{\mu} - D^{(s,t)} \boldsymbol{\alpha}\|_2^2 +  \lambda_{\mu}\boldsymbol{\mu}' L^{(t)}\boldsymbol{\mu} + \lambda_{\alpha}\boldsymbol{\alpha}' L^{(s,t)}\boldsymbol{\alpha}, \label{eq:Ridge}
\end{equation}
where $\lambda_{\mu} = \sigma^2/\sigma_{\mu}^2$ and $\lambda_{\alpha} = \sigma^2/\sigma_{\alpha}^2$ control the strength of temporal and spatio-temporal smoothing. The quadratic forms penalize differences in index values along graph edges (time and space), ensuring smooth index paths while allowing data-driven deviations where transaction evidence is strong.

\paragraph{Aggregation to SA4 and city levels.}
The SA2 indexes serve two purposes. First, we use them for exploratory PCA (Section~\ref{sec:pca}) to identify interpretable factor proxies replicating \citet{sijp_francke_ree2024}. Second, we aggregate them to construct the SA4-level and major city-level indexes used in our factor model estimation (Sections~\ref{sec:model_selection}–\ref{sec:results}). Aggregation weights are proportional to number of houses involved in sales during January 2015–January 2020 (a sub-period of our data window, selected to balance stability in transaction volumes with proximity to current market conditions), ensuring that larger markets contribute appropriately to regional composites.



\section{ARIMA parameter estimates for factor spreads}
\label{app:arima_factors}

We report ARIMA parameter estimates for the Mining and Lifestyle factor spreads. These models are used to estimate the future uncertainty in the factors themselves.
Each series is estimated in \emph{levels} via SARIMAX with an intercept, no seasonals, and stationarity/invertibility enforced. 
Orders are chosen by an AICc grid search over $p\in\{0,1,2,3\}$, $q\in\{0,1,2\}$, $d\in\{0,1\}$ with a parsimony rule: among models within $\Delta$AICc $\le 2$ at a given $d$, we pick the smallest $p{+}q$; we prefer $d{=}0$ unless the best $d{=}1$ improves AICc by $\ge 10$. 
Under this procedure, both factors select ARIMA$(2,0,1)$; residual Ljung--Box $p$-values at lags 12 and 24 are high (see Table~\ref{table:factor_funnels}).

\begin{table}[H]
  \centering
  \small
\begin{tabular}{lrrrrrrr}
\toprule
Factor & ARIMA$(p,d,q)$ & Intercept & $\phi_1$ & $\phi_2$ & $\theta_1$ & $\sigma$ (innov) & $\sigma^2$ \\
\midrule
Mining & (2,0,1) & 0.0000 & 1.9320 & -0.9340 & -0.3980 & 0.0070 & 4.485e-05 \\
Lifestyle & (2,0,1) & 0.0000 & 1.8960 & -0.8980 & -0.3090 & 0.0040 & 1.706e-05 \\
\bottomrule
\end{tabular}
  \caption{ARIMA$(2,0,1)$ estimates for factor spreads. $\sigma$ (innov) is the innovation s.d.; $\sigma^2$ its variance.}
  \label{tab:arima_params_factors}
\end{table}

For AR(2), the characteristic polynomial is $\Phi(z)=1-\phi_1 z-\phi_2 z^2$. 
Stationarity holds when the roots of $\Phi(z)=0$ lie outside the unit circle (ensured in estimation). 
With $\phi_2<0$, near-boundary values ($\phi_1\approx 2\sqrt{-\phi_2}$) correspond to highly persistent, gently \emph{damped} cycles. 
The MA(1) term $\theta_1<0$ absorbs short-run serial correlation, and the intercept is near zero because the factor series are mean-centered by construction. 
Note that $\sigma$ (innov) pertains to the one-step innovation in the state-space representation and is distinct from the 10-year forecast dispersion used in the scenario fans.

\section{Stationarity of the Mining Factor}
\label{app:mining_regimes}

In section ~\ref{sec:model_selection} the Mining factor is modelled as a persistent but stationary cycle using an ARIMA(2,0,1) specification. However, when the Augmented Dickey–Fuller (ADF) test is applied to the raw series, the unit-root null is not rejected (Table \ref{tab:adf_mining}), which appears inconsistent with the ARIMA modelling. This discrepancy arises because the series contains clear structural breaks associated with the mining boom and subsequent decline. ADF tests have low power in the presence of such level shifts and medium-term cycles, and therefore often classify a regime-shift process as non-stationary even when it is mean-reverting within segments. This appendix addresses this by identifying the break dates, analysing stationarity within each regime, and re-estimating the Mining-factor ARIMA model with regime dummies to assess how these breaks affect forecastability.

\begin{table}[H]
  \centering
  \small
  \begin{tabular}{lrr}
    \toprule
    \textbf{Series / Regime} & \textbf{ADF statistic} & \textbf{p-value} \\
    \midrule
    Raw factor (Mining)       & -1.9538 & 0.3072 \\
    First difference          & -3.9752 & 0.0015 \\
    \midrule
    Regime 1: 2000--2005      &  0.2520 & 0.9750 \\
    Regime 2: 2005--2015      & -0.7324 & 0.8382 \\
    Regime 3: 2015--2016      &  2.5719 & 0.9991 \\
    Regime 4: 2016--2020      & -2.6147 & 0.0900 \\
    \bottomrule
  \end{tabular}
  \caption{Augmented Dickey--Fuller (ADF) statistics for the Mining factor: raw series, first difference, and by structural-break regimes. The raw series does not reject the unit-root null, whereas the first difference is stationary. Regime-level tests reflect the limited power of the ADF under breaks and short samples.}
  \label{tab:adf_mining}
\end{table}

\begin{figure}[H]
  \centering
  \includegraphics[width=\textwidth]{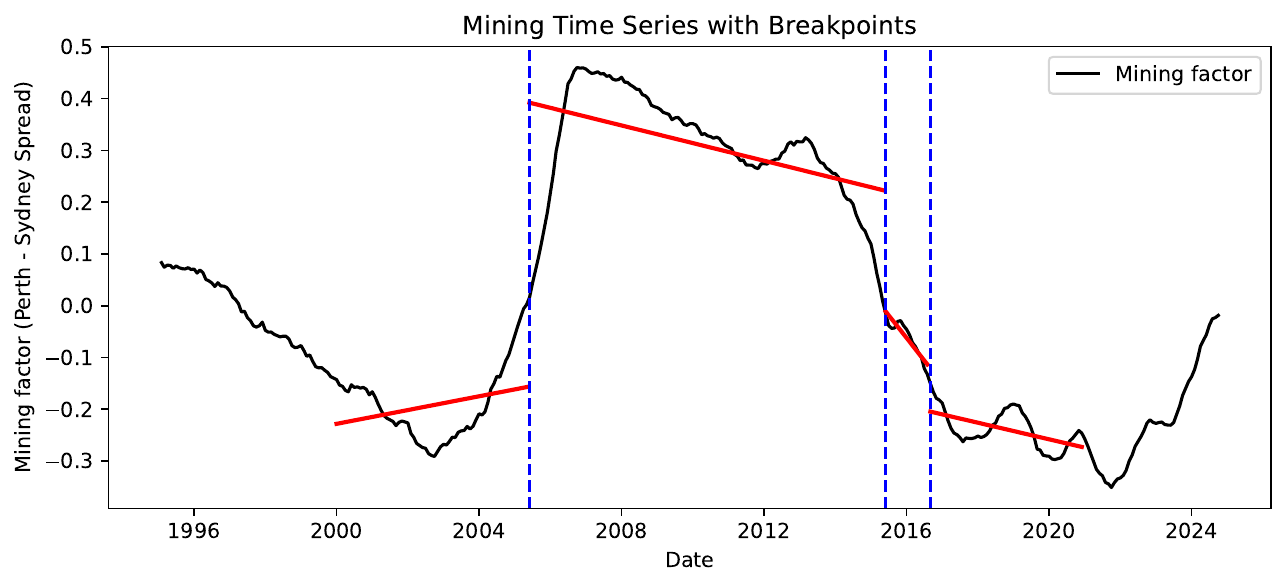}
  \caption{Breakpoint detection for the Mining factor (2000--2020) using binary segmentation.}
  \label{fig:mining_trend_breakpoints}
\end{figure}

To characterise the structural shifts in the Mining factor, we apply the ruptures algorithm to the monthly series over the 2000 to 2020 window, as this is the period that fully contains the mining boom, its peak, and the subsequent decline, and therefore captures all economically relevant regime shifts in the factor. The algorithm identifies three breakpoints that align closely with the established chronology of the Australian mining boom: an initial upward shift in mid-2005 marking the start of the boom, the peak-to-decline transition around mid-2015, and a short transitional phase in 2015 to 2016 preceding the stabilised post-2016 period captured in the 2000 to 2020 window. Figure~\ref{fig:mining_trend_breakpoints} illustrates the detected breakpoints and the fitted linear trends within each regime. These regimes correspond to those analysed in Table~\ref{tab:adf_mining}, where ADF tests are conducted for each segment.

To evaluate how structural breaks affect the forecastability of the Mining factor, we re-estimate the ARIMA model used in (Section~\ref{sec:model_selection}) but now include the three regime indicators: boom, decline, and recovery, as exogenous regressors. This model is not intended to replace the main ARIMAX used for city forecasts, rather it serves as an auxiliary check on the factor itself by explicitly controlling for the level shifts.

\begin{figure}[H]
  \centering
  \includegraphics[width=\textwidth]{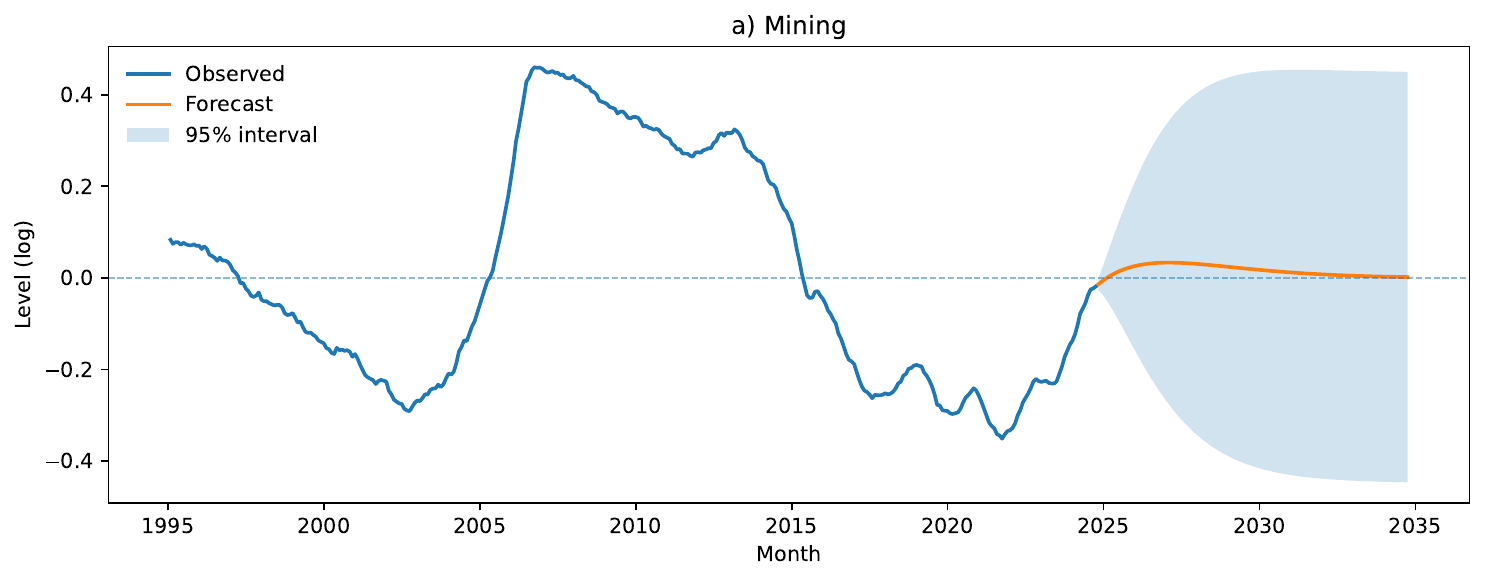}
  \caption{Ten-year forecast funnel for the Mining factor using an ARIMA(2,0,1) model with regime dummies. Accounting for the structural breaks produces a noticeably narrower $95\%$ prediction band compared with the unconditional model.}
  \label{fig:mining_break_funnel}
\end{figure}

The resulting ARIMA(2,0,1) specification remains highly persistent but clearly mean-reverting, and the inclusion of regime dummies reduces the long-horizon uncertainty relative to the unconditional model. Figure~\ref{fig:mining_break_funnel} plots the 10-year ahead forecast, showing that the 95\% prediction intervals widen more slowly and remain noticeably narrower than in the model fitted without break controls. Quantitatively, the regime-adjusted model yields a ten-year-ahead standard deviation of approximately 0.23 (compared with about 0.27 for the unconditional ARIMA), and a multiplicative 95\% prediction band of roughly 1.57 (versus around 1.70 previously). This behaviour is expected. Once the structural breaks are accounted for, the remaining dynamics are dominated by cyclical, mean-reverting fluctuations rather than long-run drift, leading to tighter and more stable long-term forecasts.

\section{Confidence Intervals}
\label{app:ci}

For completeness, we report full-sample ARIMAX estimates of the regional factor loadings and their 95\% confidence intervals in Table~\ref{tab:ci}.

\begin{table}[H]
  \centering
  \small
\begin{tabular}{lrrrrrrrrr}
\toprule
Region & $\beta_r$ & LCI & UCI & $\lambda_r$ & LCI & UCI & $\gamma_r$ & LCI & UCI \\
\midrule
Sydney & 1.01 & 0.97 & 1.05 & -0.43 & -0.47 & -0.39 & -0.15 & -0.22 & -0.07 \\
Melbourne & 1.22 & 1.20 & 1.25 & -0.22 & -0.25 & -0.18 & -0.65 & -0.71 & -0.60 \\
Brisbane & 1.00 & 0.97 & 1.03 & 0.10 & 0.06 & 0.14 & 0.37 & 0.29 & 0.45 \\
Adelaide & 0.97 & 0.90 & 1.04 & 0.06 & -0.01 & 0.12 & 0.10 & 0.01 & 0.19 \\
Perth & 0.95 & 0.91 & 0.99 & 0.59 & 0.55 & 0.63 & -0.12 & -0.19 & -0.05 \\
Hobart & 0.96 & 0.79 & 1.12 & 0.05 & -0.03 & 0.13 & 0.61 & 0.42 & 0.79 \\
Darwin & 0.73 & 0.48 & 0.99 & 0.12 & -0.00 & 0.23 & -0.21 & -0.43 & 0.01 \\
ACT & 1.00 & 0.94 & 1.05 & -0.10 & -0.19 & -0.01 & 0.08 & -0.05 & 0.22 \\
Rest Of VIC & 0.96 & 0.89 & 1.03 & 0.00 & -0.05 & 0.06 & 0.09 & 0.00 & 0.18 \\
Rest Of NSW & 0.92 & 0.89 & 0.95 & -0.04 & -0.08 & -0.01 & 0.38 & 0.33 & 0.44 \\
Rest Of QLD & 0.85 & 0.82 & 0.88 & 0.21 & 0.18 & 0.25 & 0.46 & 0.40 & 0.52 \\
Rest Of SA & 0.81 & 0.75 & 0.86 & 0.16 & 0.10 & 0.23 & 0.23 & 0.14 & 0.31 \\
Rest Of WA & 0.76 & 0.63 & 0.88 & 0.27 & 0.18 & 0.36 & -0.06 & -0.21 & 0.09 \\
Rest Of Tas. & 0.92 & 0.80 & 1.04 & 0.13 & 0.04 & 0.22 & 0.55 & 0.41 & 0.70 \\
\bottomrule
\end{tabular}
 \caption{Full-sample ARIMAX estimates of regional factor loadings and 95\% confidence intervals. The coefficients $\beta_r$, $\lambda_r$ and $\gamma_r$ are the loadings on the Market, Mining and Lifestyle factors respectively, obtained from a single ARIMAX specification estimated on the full 1995--2024 sample. LCI and UCI denote the lower and upper bounds of the 95\% confidence intervals based on regression standard errors. Coefficients values will differ somewhat from Table~\ref{table:factor_coefs}, as that table shows expanding window averages.}

  \label{tab:ci}
\end{table}

\end{document}